\documentclass[a4paper,12pt]{article}

\usepackage{amssymb,latexsym,amsmath}

\usepackage{fullpage}
\usepackage{nicefrac}
\usepackage{mathrsfs}
\usepackage{slashed}
\usepackage{bbold}
\usepackage{booktabs}
\usepackage{color,colortbl}

\usepackage{cite}

\makeatletter \@addtoreset{equation}{section}
\makeatother

\usepackage[yyyymmdd,hhmmss]{datetime}

\usepackage{xspace}
\usepackage{bbm}
\usepackage{tensind}
\tensordelimiter{`}

\usepackage[hidelinks]{hyperref}

\makeatletter
\g@addto@macro\bfseries{\boldmath}
\makeatother

\begin{document}

\begin{titlepage}
	\thispagestyle{empty}

	\hfill QMUL-PH-19-06

	\vspace*{25pt}

	\begin{center}
		
	   	{ \LARGE{\bf Uplifts of maximal supergravities and \\[3mm]
		transitions to non-geometric vacua}}

		\vspace{40pt}

		{Gianguido Dall'Agata$^{1,2}$, Gianluca Inverso$^3$ and Paolo Spezzati$^4$}

		\vspace{25pt}

		{
		$^1${\it  Dipartimento di Fisica ``Galileo Galilei''\\
		Universit\`a di Padova, Via Marzolo 8, 35131 Padova, Italy}

		\vspace{15pt}

	    $^2${\it   INFN, Sezione di Padova \\
		Via Marzolo 8, 35131 Padova, Italy}
				}

		\vspace{15pt}

			$^3${\it   Centre for Research in String Theory,  \\
			School of Physics and Astronomy, Queen Mary, University of London, \\
			327 Mile End Road, London, E1 4NS, United Kingdom \\
			}

		\vspace{15pt}

			$^4${\it   SISSA International School for Advanced Studies and INFN Trieste, \\
			Via Bonomea 265, 34136, Trieste, Italy
			}

		\vspace{40pt}

		{ABSTRACT}

		%
	\end{center}

	We describe a new procedure to obtain consistent backgrounds that uplift vacua and deformations of various maximal gauged supergravities by taking a known solution and performing singular limits along the moduli space of the corresponding 4-dimensional theory.
	We then apply this procedure to the $S^3 \times H^{2,2}$ background that provides the uplift of 4-dimensional Minkowski vacua of maximal supergravity with gauge group [SO(4) $\times$ SO(2,2)] $\ltimes \mathbb{R}^{16}$.
	We find that the newly generated vacua are generally only locally geometric and correspond to asymmetric orbifolds, $Q$-flux backgrounds or combinations thereof.
	We also provide the uplift to eleven dimensions of all the four-parameter Cremmer--Scherk--Schwarz gaugings.

	\vspace{10pt}

\end{titlepage}

\baselineskip 6 mm

\renewcommand{\arraystretch}{1.2}

\tableofcontents


\definecolor{shaded}{gray}{.85}

\allowdisplaybreaks

\section{Introduction} 
\label{sec:introduction}

Double field theory (DFT) \cite{Siegel:1993xq, Siegel:1993th, Hull:2009mi, Hohm:2010jy}
 and Exceptional Field Theory (ExFT) \cite{Hull:2007zu,Pacheco:2008ps,Hillmann:2009pp,Berman:2010is,Berman:2011pe,Coimbra:2011ky,Berman:2012vc,Coimbra:2012af,Hohm:2013vpa,Hohm:2013uia,Hohm:2014fxa} have been very effective tools to understand and clarify the relation between lower dimensional (gauged) supergravities and their 10 or 11-dimensional counterpart.
This is clearly important in the context of better understanding the possible landscape of effective theories that have a string uplift.
It is also instrumental to the identification of new consistent truncations of string theory reductions to lower dimensions in order to gain control on the vacua of the theory, their deformations and their relations, which are often evident in DFT and ExFT because of their manifest duality covariance.
A point that is especially important is the realization that one can generalize Scherk--Schwarz (SS) reductions \cite{Scherk:1978ta,Scherk:1979zr} to DFT and ExFT in a way that allows to make contact between supergravity gaugings and 10 and 11-dimensional backgrounds in an explicit fashion \cite{Aldazabal:2010ef,Berman:2012uy,Musaev:2013rq,Aldazabal:2013mya,Lee:2014mla,Hohm:2014qga}.

In this work we focus on $D=4$ maximal supergravity and its Minkowski vacua.
It was noted that there is a large class of Minkowski vacua of $D=4$ gauged maximal supergravities that are connected by singular limits along their moduli spaces \cite{Catino:2013ppa}.
In particular, there is a gauging with gauge group $[{\rm SO}(4)\times {\rm SO}(2,2)]\ltimes \mathbb{R}^{16}$ that is known to derive from the reduction of type IIB string theory on the $S^3 \times H^{2,2}$ manifold \cite{Inverso:2016eet,Baguet:2015iou,Malek:2017cle}.
We therefore decided to analyze singular limits along the moduli space of this 4-dimensional model from the higher-dimensional perspective, by means of DFT, and provide a procedure to construct the corresponding backgrounds and deformations.
In the process, we give a general procedure for these singular deformations, which goes beyond the specific examples discussed in this work.

The first result that we are going to present is a general procedure by which singular limits of a gauged supergravity that has a generalized SS uplift to 10 or 11 dimensions yield novel gauged supergravities that also have a similar uplift. 
By means of our constructive procedure, we find that the twist matrix of the limit model can be obtained from the original twist matrix.
This makes everything very explicit and allows us to write for each limit the local functions describing the metric, the dilaton and all the other form fields present in the theory.

As a working example, we took singular limits along the moduli space of Minkowski vacua of the SO(4) $\times$ SO(2,2) gauging and its siblings identified in \cite{Catino:2013ppa}.
We discovered that all such models result in new Minkowski$_4$ \textit{solutions} of 10-dimensional string theory with an internal space some type of $T$-fold \cite{Dabholkar:2002sy,Hellerman:2002ax,Flournoy:2004vn,Hull:2004in,Dabholkar:2005ve} admitting an interpretation as asymmetric orbifolds, $Q$-flux backgrounds and combinations of the two.

We close our work with a presentation of the 11-dimensional uplift of all CSS gaugings \cite{Cremmer:1979uq}.
A three-parameter subclass of these models was originally found by Scherk and Schwarz by dimensional reduction of eleven-dimensional supergravity\cite{Scherk:1978ta}, however the origin of the fourth parameter was still missing and we now close this gap.


\section{The moduli space of Minkowski vacua in $N=8$ supergravity} 
\label{sec:the_moduli_space_of_minkowski_vacua_in_n_8_supergravity}

Gauged $N=8$ supergravity models are completely fixed once the embedding tensor $\Theta$ is specified \cite{deWit:2002vt,deWit:2007kvg}.
The $\Theta_M{}^\alpha$ tensor declares which generators $t_{\alpha}$ of the E$_{7(7)}$ duality group are made local by means of the vector fields (and their duals) $A_\mu^M$ present in the theory:
\begin{equation}
	D_\mu = \partial_\mu - A_\mu^M X_M, \qquad X_M = \Theta_M{}^\alpha t_{\alpha}.
\end{equation}
Different gauge groups $G_{gauge}$ have different embedding tensors, but sometimes also the same group $G_{gauge}$ can be embedded in different ways in the duality group and the embedding tensor provides the (different) resulting lagrangians.
Also, the analysis of invariant values of quantities constructed in terms of the embedding tensor discriminates equivalent and inequivalent models \cite{DallAgata:2012mfj,DallAgata:2014tph}.

A very important term in the Lagrangian is the scalar potential, which is a quadratic function of the embedding tensor and which determines the vacua of the theory and their residual symmetry group $G_{res} \subset G_{gauge}$.
For maximal supergravity the scalar manifold is the quotient E$_{7(7)}/$SU(8) and therefore one can perform the minimization of the scalar potential required to establish the vacua for a given model by using directly the embedding tensor $\Theta$ and scanning the values satisfying at the same time the extremum conditions and the consistency conditions coming from the gauging procedure \cite{DallAgata:2011aa}.
Such trick allowes to determine systematically a large part of the spectrum of vacua of maximal supergravity with various numbers of supersymmetries and values of the cosmological constant \cite{DallAgata:2011aa,DallAgata:2012mfj, Borghese:2012qm,Borghese:2012zs,Borghese:2013dja,Catino:2013ppa,Gallerati:2014xra}.
In particular, in the analysis presented in \cite{Catino:2013ppa}, many new marginally stable Minkowski vacua were found for different gauge groups $G$ and embeddings $\Theta$, preserving $N=0,2,4$ and 6 supersymmetries.
This analysis vastly generalized the sparse set of previously known models having Minkowski vacua \cite{Cremmer:1979uq,Hull:2002cv}.

An especially interesting aspect is given by the fact that all these vacua are obtained by contractions of a single gauged supergravity model with $G_{gauge} =$ SO$^*(8)$ and a specific dyonic choice of its embedding in the duality group \cite{Catino:2013ppa}.
This observation is at the basis of the present work, because the contractions leading to new models with Minkowski vacua could be interpreted as deformations of the background geometry leading to the original model if one could uplift the original vacuum to 10- or 11-dimensional supergravity.

In detail, in the SO$^*(8)$ gauged supergravity model analyzed in \cite{Catino:2013ppa}, there are 48 massless scalar fields at the maximally symmetric point.
Only 20 of them are real moduli fields, because the others are Goldstones or would-be Goldstone fields.
However, after giving an arbitrary vacuum expectation value to any of these fields one breaks $G_{res} =$ SO(6) $\times$ SO(2), maintaining always at least 6 massless fields.
These are the real moduli of the model and they parameterize a $\left[\frac{\rm SU(1,1)}{\rm U(1)}\right]^3$ scalar manifold, with fields labelled $e_i$ and $x_i$, for $i=1,2,3$, in \cite{Catino:2013ppa}.
These moduli can be used to obtain new gaugings following the procedure we now detail.

The approach of \cite{Catino:2013ppa}, which we now review, was already used in \cite{Hull:1984qz,Hull:1984yy,Hull:2002cv} to produce the CSO$(p,q,r)$ and CSO$^*(2p,2q)$ gaugings, but it was then applied in more generality in \cite{Catino:2013ppa} by employing the embedding tensor formalism, as in \cite{Fischbacher:2003yw}.
The idea is to introduce a one-parameter deformation of the gauging, associated with the action on $\Theta_M{}^\alpha$ of some (non-compact) $\rm E_{7(7)}$ duality in order to fulfill the consistency constraints, and then take a singular limit to produce an inequivalent gauging.
Suppose we parametrize a geodesic in E$_{7(7)}/{\rm SU}(8)$ as $G(\xi) = e^{t \log \xi}$, where $\xi\in \mathbb R^*_+$, for some generator $t=t^T$ of the coset space. 
Then the boundary of E$_{7(7)}/\rm SU(8)$ is reached at $\xi \rightarrow 0, +\infty$.
Starting from a gauged model with embedding tensor $\Theta$, we can then define a $\xi$-dependent embedding tensor by the appropriate action of the fundamental and adjoint representations of $G(\xi)$ on $\Theta$:
\begin{equation}
\Theta(\xi)_M{}^\alpha\equiv \left[G(\xi)\Theta\right]_M{}^\alpha \equiv G(\xi)_M{}^N \Theta_{M}{}^\beta G(\xi)_\beta{}^\alpha.
\end{equation}
For $\xi\in \mathbb R^*_+$, $\Theta(\xi)$ still gauges a group isomorphic to the one defined by $\Theta$. 
However, taking the limit $\xi\rightarrow0$, $G(\xi)$ becomes singular and $\Theta(\xi)$ diverges because some of its entries depend on negative powers of $\xi$. 
We can still obtain a finite embedding tensor if we pair the previous limit with a rescaling of the coupling constant 
\begin{equation}
	g\rightarrow g' \xi^p,
\end{equation} 
where $p$ is chosen according to the most singular entries of $\Theta(\xi)$, proportional to $\xi^{-p}$. 
Hence we define:

\begin{equation}\label{contraction general definition}
g'_{\rm gauge} \Theta_{\rm contr.} \equiv \lim_{\xi\rightarrow0} \left[g'\, \xi^p\, \Theta(\xi)\right].
\end{equation}
Now $\Theta_{\rm contr}$ defines a $\rm G_{gauge}$ which is generally not isomorphic to the original one.

Since the action of $G(\xi)$ on $\Theta_{\rm contr.}$ commutes with the limit in \eqref{contraction general definition},  we see that the contracted embedding tensor has a grading $-p$ with respect to the generator along which we performed the contraction:
\begin{equation}
G(\xi) \Theta_{\rm contr.} = \xi^{-p} \Theta_{\rm contr.}.
\end{equation}
This implies that further contractions along $t$ do not produce new gaugings.

The contraction procedure defined in \eqref{contraction general definition} can be applied for any generic direction in E$_{7(7)}/{\rm SU}(8)$, but we now focus on singular limits along the moduli space $\left[{\rm SU}(1,1)/{\rm U}(1)\right]^3$, hence preserving not only the embedding tensor constraints but also the vacuum condition.
This means that the parameter $\xi$ in the following is going to be identified with one of the moduli.

In order to perform these contractions, we start from the embedding tensor $\Theta_0^{\mathfrak{so}^*(8)}$ of SO$^*(8)$, defined at the maximally symmetric point where $G_{\rm res} = \rm SU(4)\times U(1)$, and act on it with the transformations:
\begin{align}
X_i \equiv \exp(\ell_i\,\log x_i),\quad E_i \equiv \exp(\lambda_i\,\log e_i),\quad i=1,2,3,
\end{align}
where $\ell_i = \ell^T_i,\ \lambda_i=\lambda^T_i$ generate each ${\rm SU}(1,1)/{\rm U}(1)$ factor.
The three factors commute with each other, but since $[\ell_i, \lambda_i] \neq0$, we need to fix the order in which they act on $\Theta_0^{\mathfrak{so}^*(8)}$:
\begin{equation}
\Theta^{\mathfrak{so}^*(8)}(x_i,\,e_i)\equiv\prod_{i=1}^3 (X_i\, E_i) \Theta_0^{\mathfrak{so}^*(8)}.
\end{equation}
Of course, any other ordering or parametrization of the coset space is equivalent up to a change of coordinates.

The combinations of the various limits that we can now perform in the $x_i$ and $e_i$ directions give rise to several models with Minkowski vacua and spontaneously broken supersymmetry.
We summarize the resulting contracted models in Table \ref{tab:contractions}.
Taking the singular limits in different orders always reproduces one of the gauge groups in Table~\ref{tab:contractions}, and the same mass spectra are obtained up to a reordering of the moduli. 

\begin{table}\small
\begin{center}
\begin{tabular}{r|cccc}
& & $x_1\rightarrow0$ & $x_1, x_2\rightarrow0$ & $x_1, x_2, x_3\rightarrow0$\\
\hline\\[-.8em]
& SO$^*(8)$ & \cellcolor{shaded} $[{\rm SO}(4)\times {\rm SO}(2,2)]\ltimes \mathbb{R}^{16}$ & \cellcolor{shaded}$[{\rm U}(1)^2\ltimes N^{26}]_{\mathcal N=0}$ & \cellcolor{shaded}${\rm CSS}_{\mathcal N=0}$ \\[.2em]
$e_3\rightarrow0$ & ${\rm CSO}^*(4,4)$ & \cellcolor{shaded}$[{\rm U}(1)^2 \ltimes N^{20}]_{\mathcal N=4}$ & \cellcolor{shaded}${\rm CSS}_{\mathcal N=4}$ &\cellcolor{shaded} ${\rm CSS}_{\mathcal N=4}$ \\[.2em]
$e_2, e_3\rightarrow0$ & ${\rm CSS}_{\mathcal N=6}$ &\cellcolor{shaded} ${\rm CSS}_{\mathcal N=6}$ & \cellcolor{shaded}${\rm CSS}_{\mathcal N=6}$ & \cellcolor{shaded}${\rm CSS}_{\mathcal N=6}$ \\[.2em]
$e_1^{-1}, e_2, e_3\rightarrow0$ & ${\rm CSO}^*(6,2)$ & $[{\rm SO}^*(4)\times {\rm U}(1)]\ltimes {N^{20}}$ & $[{\rm U}(1)^2\ltimes N^{24}]_{\mathcal N=2}$ & ${\rm CSS}_{\mathcal N=2}$ 
\end{tabular}
\end{center}
\caption{Contractions along the moduli space of the SO$^*(8)$. 
The shaded models are those for which we find uplifts by taking singular limits of a $S^3\times H^{2,2}$ internal geometry.
The full class of CSS gaugings is uplifted to eleven dimensions in section~\ref{sec:other_uplifts}.
}\label{tab:contractions}
\end{table}

As proved in \cite{Inverso:2017lrz}, the SO$^*(8)$ model described here does not admit a (globally or locally) geometric uplift. 
As we will describe in the following, we will use instead the $S^3 \times {\cal H}^{2,2}$ reduction of \cite{Baguet:2015iou,Inverso:2016eet,Malek:2017cle} as the starting manifold, which corresponds to the $[{\rm SO}(4) \times {\rm SO}(2,2)]\ltimes \mathbb{R}^{16}$ gauging of maximal supergravity first constructed in \cite{DallAgata:2011aa}.
We will then identify the geometric deformations corresponding to the moduli of the reduced theory.
From the 4-dimensional point of view this corresponds to the contractions denoted in Table~\ref{tab:contractions} by a shaded background.

As explained in \cite{Catino:2013ppa}, contractions along a modulus $x_i$ generate families of inequivalent gaugings parameterised by the vev assigned to $e_i$  before contraction. 
For instance, the [SO(4) $\times$ SO(2,2)] $\ltimes {\mathbb R}^{16}$ entry in Table~\ref{tab:contractions} really corresponds to a one parameter family of gauged supergravities parameterised by the value assigned to $e_1$ in the SO$^*(8)$ model, before taking the $x_1\to0$ limit.
Instead, changing the value of $e_1$ after the limit will only rescale the embedding tensor by an overall factor.
It is important to stress that the only member of this family of [SO(4) $\times$ SO(2,2)] $\ltimes {\mathbb R}^{16}$ gaugings that has been uplifted in \cite{Baguet:2015iou,Inverso:2016eet,Malek:2017cle} is the one corresponding to leaving $e_1 = 1$ in the SO$^*(8)$ model before taking $x_1\to 0$.
As a result, only a subset of the models appearing on Table 1 can be reached starting from this one [SO(4) $\times$ SO(2,2)] $\ltimes {\mathbb R}^{16}$ gauging and in particular the last row cannot be reached.

Finally, it is interesting to note that all the relevant moduli of the [SO(4) $\times$ SO(2,2)] $\ltimes {\mathbb R}^{16}$ model are already contained in an $N=4$ truncation and that therefore we can obtain all the desired deformations in a simplified setup.

\subsection{Potential and moduli of half-maximal supergravity} 
\label{sub:potential_and_moduli_of_half_maximal_supergravity}

The relevant truncation of half-maximal supergravity has a potential dependent on $37$ scalar fields, which can be determined by the structure constants $X_{AB}{}^C$, truncated to the SO(6,6) generators of the scalar $\sigma$-model ${\mathbb R} \times {\rm SO}(6,6)/[{\rm SO}(6)\times {\rm SO}(6)]$:
\begin{equation}
	V(x) = \frac{1}{12} \,e^{2 \varphi(x)} X_{AB}{}^C X_{DE}{}^F M^{AD}(x) \left(M^{BE}(x) M_{CF}(x) + 3\, \delta_B^E \delta_F^C\right).
\end{equation}
Here $M_{AB}(x)$ is an SO(6,6) matrix satisfying $M^T \eta M = \eta$, where $\eta = \left(\begin{array}{cc}
{\mathbb 0}_6 & {\mathbb 1}_6 \\
{\mathbb 1}_6 & {\mathbb 0}_6 
\end{array}\right)$ is the SO(6,6) invariant metric.
It is immediately clear that such potential can only produce Minkowski vacua because of the overall dependence on the dilaton $\varphi(x)$.

In order to assess the structure of the vacua we will find, it is useful to give an explicit expression for the SO(6,6) generators in this basis.
We will split the generators in 5 sets, labeled by indices $i,j=1,\ldots,6$.
The compact generators are antisymmetric and either block diagonal
\begin{equation}
	(A^{ij})_A{}^B = \frac12 \left(\delta_A^i \delta^{jB} - \delta^j_A \delta^{iB}\right) + \frac12  \left(\delta_A^{i+6} \delta^{j+6\,B} - \delta^{j+6}_A \delta^{i+6\,B}\right),
\end{equation}
or block off-diagonal
\begin{equation}
	(C^{ij})_A{}^B = \frac12 \left(\delta_A^i \eta^{jB} - \delta^j_A \eta^{iB}\right) + \frac12  \left(\delta_A^{i+6} \eta^{j+6\,B} - \delta^{j+6}_A \eta^{i+6\,B}\right).
\end{equation}
The non-compact generators are symmetric and block diagonal
\begin{align}
	(d^i)_A{}^B &= \frac{1}{\sqrt2} \left(\delta^i_A \delta^{iB} - \delta^{i+6}_A \delta^{i+6\,B}\right), \\[2mm]
	(S^{ij})_A{}^B &= \frac12 \left(\delta_A^i \delta^{jB} + \delta^j_A \delta^{iB}\right) - \frac12  \left(\delta_A^{i+6} \delta^{j+6\,B} + \delta^{j+6}_A \delta^{i+6\,B}\right),
\end{align}
or block off-diagonal
\begin{equation}
	(T^{ij})_A{}^B = \frac12 \left(\delta_A^i \eta^{jB} - \delta^j_A \eta^{iB}\right) - \frac12  \left(\delta_A^{i+6} \eta^{j+6\,B} - \delta^{j+6}_A \eta^{i+6\,B}\right).
\end{equation}
Altogether we have $t_{\alpha} = \{A,C,d,S,T\}$ satisfying $tr(t_{\alpha}t_{\beta}) = \mp \, \delta_{\alpha \beta}$, depending on their being compact or not.

The truncation of the $N=8$ model with $G_{gauge} = {\rm SO}(4) \times {\rm SO}(2,2)$ to $N=4$ gives a gauged supergravity model with $G_{gauge} = SO(4) \times SO(2,2)$.
The corresponding gauging follows from fixing the gauge generators $X_{AB}{}^C = \Theta_A{}^\alpha (t_{\alpha})_B{}^C$ by choosing
\begin{equation}\label{Xdef}
	\begin{array}{rl}
	X_1 &= \sqrt2\, C^{23}, \quad X_2 = - \sqrt2\, C^{13}, \quad X_3 = \sqrt2\, C^{12}, \\[2mm]
	X_4 &= -\sqrt2\, T^{56}, \quad X_5 = \sqrt2\, T^{46}, \quad X_6 = -\sqrt2\, C^{45}, \\[2mm]
	X_7 &= \sqrt2\, A^{23}, \quad X_8 = - \sqrt2\, A^{13}, \quad X_9 = \sqrt2\, A^{12}, \\[2mm]
	X_{10} &= \sqrt2\, S^{56}, \quad X_{11} = - \sqrt2\, S^{46}, \quad X_{12} = \sqrt2\, A^{45}, 
	\end{array}
\end{equation}
so that the corresponding $G_{gauge} \times G_{gauge}$ invariant Cartan--Killing form has entries
\begin{equation}
	\kappa_{AB} = \frac12\, X_{AC}{}^D X_{BD}{}^C,
\end{equation}
normalized to $\pm 1$.
The resulting potential has a critical point where $G_{res} = {\rm SU}(2) \times {\rm SU}(2) \times {\rm U}(1) \times {\rm U}(1)$ and the mass spectrum can be arranged according to the $G_{res}$ representations
\begin{equation}
\begin{array}{c|c|l}
{\rm mass} & {\rm irrep} & {\rm generators} \\\hline
0 & (3,3)_{0,0} & S^{12}, S^{13}, S^{23}, T^{12}, T^{13}, T^{23}, d^1, d^2, d^3 \\[2mm]
0 & (1,1)_{0,0} & d^6 \\[2mm]
0 & (1,3)_{0,0} + (3,1)_{0,0} & S^{16}, S^{26}, S^{36}, T^{16}, T^{26}, T^{36}\\[2mm]
2 & (1,3)_{\pm,\mp} + (3,1)_{\mp,\pm}  &  S^{14}, S^{15}, S^{24}, S^{25}, S^{35}, T^{14}, T^{15}, T^{24}, T^{25}, T^{35}\\[2mm]
2 & 2 (1,1)_{\pm,\mp}  & S^{46}, S^{56}, T^{46}, T^{56} \\[2mm]
4 & (1,1)_{0,\pm2} + (1,1)_{\pm2,0} & S^{45}, T^{45}, d^4, d^5
\end{array}
\end{equation}
where we arbitrarily fixed the overall scale, determined by the dilaton, which is also an overall modulus in the potential.
Among the moduli we recognize the [SU(1,1)/U(1)]$^2$ factor, generated by $\lambda_2 = S^{36}$, ${\ell}_2 = d^3 - d^6$ and $\lambda_3 = T^{36}$, $\ell_3 = d^3 + d^6$.
The contractions leading to the deformed vacua will follow by introducing the moduli dependence in the embedding tensor and taking their limit to the boundary.



\section{Ten dimensional origin} 
\label{sec:ten_dimensional_origin}

\subsection{Extended field theories and generalized Scherk--Schwarz reductions} 
\label{sub:extended_field_theories}

Double and exceptional field theories (DFT and ExFT) \cite{Siegel:1993xq, Siegel:1993th, Hull:2009mi,Hohm:2010jy,Hohm:2013vpa,Hohm:2013uia,Hohm:2014fxa,Abzalov:2015ega,Musaev:2015ces,Berman:2015rcc,Bossard:2017aae,Bossard:2018utw} encode the 10- and 11-dimensional supergravity theories in a framework formally covariant under O$(d,d) \times {\mathbb R}^+_{\Phi}$ and E$_{n(n)}$ groups, respectively. 
These groups are the global symmetries of the lower-dimensional supergravities obtained upon Kaluza--Klein reducing on tori, but crucially appear in DFT and ExFT as generalized structure groups before any truncation.
More generally, duality groups other than O$(d,d) \times {\mathbb R}^+_{\Phi}$ and E$_{n(n)}$ can be also encoded in a similar formalism (see in particular \cite{Strickland-Constable:2013xta,Ciceri:2016hup,Hohm:2017wtr,Cederwall:2017fjm}).
We shall hence denote the generic duality group ${\cal G}$.
In these frameworks, fields depend on an `external' spacetime with coordinates $x^\mu$, as well as an internal space whose coordinates $y^m$ are formally extended to $Y^M$, filling a representation $\mathbf R_{\rm v}$ of ${\cal G}$.\footnote{Double field theory can be formulated by doubling the coordinates of the entire spacetime, but for our purposes we prefer its formulation including a non-doubled external spacetime \cite{Hohm:2013nja}, which is more useful to perform dimensional consistent truncations and closely follows the structure of ExFTs.}
The theories are formally invariant under rigid ${\cal G} \times {\mathbb R}^+$ transformations, where ${\mathbb R}^+$ is the trombone symmetry that also acts on the external Einstein frame metric. 
Consistency requires to impose a `strong' or `section' constraint on the dependence of fields on the extended coordinates $Y^M$, which effectively reduces them to only depend on a set of physical internal coordinates and breaks the global ${\cal G} \times {\mathbb R}^+$ invariance.
Upon solving the section constraint, DFT and ExFT reproduce the dynamics of ten- and eleven-dimensional supergravities phrased in terms of the associated (exceptional) generalized geometries \cite{Coimbra:2011ky,Coimbra:2012af}.

The bosonic field content of DFT/ExFT is constituted by an external metric $g_{\mu\nu}(x,Y)$, scalar fields parameterizing a coset space ${\cal G}/K({\cal G})$, vector fields $A_\mu^M(x,Y)$, and so on for higher $p$-forms in other representations of the duality groups.
The gauge symmetries of DFT and ExFT along the internal space are encoded in terms of generalized vectors $\Lambda^M(x,Y)$, acting on fields by means of a generalized Lie derivative which is most easily defined by its action on another generalized vector\footnote{We shall henceforth exclude theories that require covariantly constrained gauge parameters \cite{Hohm:2014fxa,Bossard:2017aae,Cederwall:2017fjm}.} $V^M$:
\begin{equation}
{\mathbb L}_\Lambda V^M = \Lambda^N\partial_N V^M - V^N\partial_N\Lambda^M +Y^{MP}{}_{QN}\partial_P\Lambda^Q\,V^N\,.
\end{equation}
The invariant tensor $Y^{MN}{}_{PQ}$ depends on the theory and encodes the projection of the matrix $\partial_M\Lambda^N$ onto the  algebra of generators of the generalized structure group \cite{Berman:2012vc}.
Finally, closure of the generalized Lie derivative and consistency of the dynamical theory require the partial derivatives $\partial_M=\partial\,/\partial Y^M$ to satisfy the section constraint
\begin{equation}
Y^{MN}{}_{PQ} \partial_M\otimes\partial_N = 0
\end{equation}
when acting on any field or product of fields, effectively restricting the coordinate dependence of all fields on a set $(x^\mu,\,y^m)$ of physical coordinates.
There are two separate orbits of maximal solutions of the section constraint for ExFTs, corresponding to eleven dimensional and IIB supergravity respectively. 
In the case of DFT, only one maximal solution of the section constraint is available, corresponding to minimal 10-dimensional supergravity.%
\footnote{Vector couplings for heterotic supergravity can be also encoded in DFT by extending O$(d,d)$ to O$(d,d+n)$ \cite{Hohm:2011ex}, and massive IIA supergravity is encoded in ExFT through a deformation of the generalized Lie derivative \cite{Ciceri:2016dmd,Cassani:2016ncu}.}

DFT and ExFT are effective frameworks to study consistent dimensional truncations of supergravity theories.
We shall focus here on generalized Scherk--Schwarz reductions, which are determined by a twist matrix or generalized frame $\hat E_A{}^M(Y)$ (which we will always assume to satisfy the section constraint) taking values in ${\cal G}\times{\mathbb R}^+$ and defining the factorization of the internal coordinate dependence for covariant fields according to their representation under the duality group (as we will show explicitly in the next section). 
Upon factorization of the $Y^M$ dependence, the theory reduces to a lower-dimensional gauged supergravity.
Consistency of the truncation requires the twist matrix to satisfy a differential equation
\begin{equation}\label{gSS condition}
{\mathbb L}_{\hat E_A} \hat E_B{}^M = -`X_AB^C`\hat E_C{}^M\,,
\end{equation}
where $X_{AB}{}^C$ is the embedding tensor of the resulting gauged supergravity.


\subsection{$[{\rm SO}(4) \times {\rm SO}(2,2)]\ltimes \mathbb{R}^{16}$ uplift} 
\label{sub:_rm_so_4_times_rm_so_2_2_ltimes_t_16_uplift}

We now review the results of \cite{Baguet:2015iou,Inverso:2016eet,Malek:2017cle}, where the uplift of the $[{\rm SO}(4) \times {\rm SO}(2,2)]\ltimes \mathbb{R}^{16}$ gauging has been provided.
We will actually mainly focus on the half-maximal truncation and therefore on the uplift of the common sector of 10-dimensional supergravities as in \cite{Baguet:2015iou}.
The main point of the construction in \cite{Baguet:2015iou} is the fact that consistent truncations of the common sector of 10-dimensional supergravities are easily obtained and described by generalized Scherk--Schwarz reductions of DFT.
We will employ the formulation of \cite{Hohm:2013nja}, in which the SO(6,6) duality is manifest, provided the 10-dimensional degrees of freedom, the metric, the dilaton and the 2-form are combined in the SO(6,6) covariant fields $\Phi$, ${\cal B}$, ${\cal H}_{MN}$ and $A^M_\mu$, where the SO(6,6) indices split as $Y^M = \{ y^m, y_m\}$, as follows:
\begin{align}\label{10dorigin}
	A_\mu^m &= G^{mn} G_{\mu n}, \quad  A_{\mu m}  = - B_{\mu m} + A_\mu^n B_{nm}, \\[2mm]
	{\cal B}_{\mu\nu} &= B_{\mu \nu} + 2 A_{[\mu}^m B_{\nu]m} + A_{[\mu}^m A_{\nu]}^n B_{mn} + A^m_{[\mu}A_{\nu]m}, \\[2mm]
	g_{\mu\nu} &= e^{\frac{\phi}{2}}\left(G_{\mu\nu} - A_\mu^m A_\nu^n G_{mn}\right) \\[2mm]
	{\cal H}^{mn} &= e^{-\frac{\phi}{2}}G^{mn}, \qquad {\cal H}_m{}^n = e^{-\frac{\phi}{2}} G^{nk}B_{km}, \\[2mm]
	{\cal H}_{mn} &= e^{-\frac{\phi}{2}} G^{kl}B_{km}B_{ln} + e^{\frac{\phi}{2}} G_{mn}, \\[2mm]
	e^{\Phi} &= e^{\frac{\phi}{4}} \left({\rm det}G_{mn}\right)^{-1/4}.\label{10dorigfin}
\end{align}
From these fields one can write down an O(6,6) covariant action, which can be found in \cite{Hohm:2013nja}.
In order to solve the section constraint and reproduce the supergravity equations of motion, we take all fields to depend only on $x^\mu$ and $y^m$.

What is interesting for us is that it has been shown that one can obtain consistent truncations of the bosonic string sector to 4 dimensions by means of a generalized Scherk--Schwarz reduction. 
Let us discuss how this works.
In our current formulation, DFT exhibits an O$(6,6)\times{\mathbb R}_{\Phi}^+\times{\mathbb R}^+$ symmetry, where the first factor is a shift of the O(6,6) invariant dilaton $\Phi$ and the second factor is the four-dimensional trombone symmetry.
A generic twist matrix is thus valued in this group.
The general expressions for a generalized Scherk--Schwarz reduction were derived in \cite{Geissbuhler:2011mx}, see also section~4 of \cite{Ciceri:2016hup} for a derivation that includes the SL(2) part of the four-dimensional duality group, as well as trombone gaugings.
Generalized vectors have opposite weights with respect to the two scaling symmetries.
The twist matrix thus reads
\begin{equation}
\hat E(y)_A{}^M = e^{-\tau(y)}e^{d(y)} U^{-1}(y)_A{}^M\,, \quad U(y)_M{}^A\in O(6,6)\,,\quad e^{-\tau(y)}\in{\mathbb R}^+\,,\quad e^{-d(y)}\in{\mathbb R}^+_\Phi\,.
\end{equation}
There is no evident distinction between $\tau(y)$ and $d(y)$ in the twist matrix, but they play a different role in the reduction ansatz for the fields, where they appear with powers proportional to their ${\mathbb R}^+$ and ${\mathbb R}^+_\Phi$ charges.
We are only interested in twist matrices giving rise to gaugings valued in SO(6,6), which requires to identify these two factors in the twist matrix into a single function $\rho(y)$ so that we have (possibly up to a constant factor that can be reabsorbed in a redefinition of the four dimensional fields)
\begin{equation}
e^{\tau(y)} = e^{d(y)} = \rho(y)^{1/2}\,.
\end{equation}
Then, $\hat E(y)_A{}^M = U^{-1}(y)_A{}^M$ takes values in SO(6,6).

The $y$ dependence of the DFT fields is factorized according to the following ansatz
\begin{align}
	{\cal H}_{MN}(x,y) &= U_M{}^A(y) M_{AB}(x)U_N{}^B(y), \quad e^{\Phi}(x,y) = \rho(y) e^{\varphi(x)}, \\[2mm]
	{\cal A}_\mu^M(x,y) &= (U^{-1})_A{}^M(y) A_\mu^A(x), \\[2mm]
	{\cal B}_{\mu\nu}(x,y) &= b_{\mu\nu}(x), \\[2mm]
	g_{\mu\nu}(x,y) &= e^{\varphi(x)/2} g_{\mu\nu}(x),
\end{align}
where $A_\mu^A(x)$, $b_{\mu\nu}(x)$, $\varphi(x)$ and $M_{AB}(x)$ are the vector, tensor and scalar fields of the resulting 4-dimensional half-maximal supergravity.
In particular, $\varphi(x)$ is the 4-dimensional dilaton and $M_{AB}(x)$ coincides with the scalar matrix described in the previous section.

Consistency of the reduction requires to impose \eqref{gSS condition} (with $Y^{MN}{}_{PQ} = \eta^{MN}\eta_{PQ}$). 
After a bit of algebra, restricting to SO$(6,6)$ gaugings these conditions reduce to the ones exhibited in \cite{Baguet:2015iou}:
\begin{align}\label{cons1}
	\eta_{D[A} (U^{-1})_B{}^M (U^{-1})_{C]}{}^N \partial_M U_N{}^D &= f_{ABC} = {\rm const}., \\[2mm]
	\rho^{-1} \partial_M \rho & = - \frac12\, (U^{-1})_A{}^N \partial_N U_M{}^A, \label{cons2}
\end{align}
where $f_{ABC}$ determine the gauging by means of the embedding
\begin{equation}
	X_{AB}{}^C = f_{ABD}\eta^{DC}.
\end{equation}
In the special case at hand we can exploit the fact that the compactification manifold is also a product of groups and use group related quantities to construct the twist matrices $U(y)$.
This can be described by a product of matrices
\begin{equation}
	U_M{}^A = [U_0 R(\theta_2,\psi_2,\theta_3,\psi_3)]_M{}^A,
\end{equation}
where
\begin{equation}
	U_0 = \left(\begin{array}{cc}
	D & {\mathbb 0}_6 \\[2mm]
	Z & D^{-1}
	\end{array}\right), 
\end{equation}
with
\begin{equation}
	D = \sqrt2\,{\rm diag}\left\{1, 1, \tan \theta_1, 1, 1,  \tanh \psi_1\right\}
\end{equation}
and
\begin{equation}
	Z = \frac{1}{\sqrt2}\left(\begin{array}{cccccc}
	0 &  &  &  &  & \\
	 & 0 & \tan \theta_1  &  &  & \\
	& -1 & 0 &  &  & \\
	 &  &  & 0 &  & \\
  &  &  &  & 0 & \tanh \psi_1\\
   &  &  &  & - 1 & 0\\
	\end{array}\right)
\end{equation}
contains all the local information on the 10-dimensional geometry at the maximally symmetric point $M_{AB} = \delta_{AB}$, while $R$ is a rotation matrix belonging to the SO(6) $\times$ SO(6) subgroup of the duality group and has the following explicit form:
\begin{equation}\label{R}
	\begin{split}
	R =& \exp\left[-2 \theta_2 \,C^{13}+2\left(\psi_2+\frac{\pi}{2}\right)\,C^{46}\right]\,\exp\left[2 \left(\theta_3+\frac{\pi}{2}\right) A^{13}+2\left(\psi_3+\frac{\pi}{2}\right)A^{46}\right]  \\[2mm]
	&\, \cdot \exp\left[\pi\left(A^{23}+A^{56}\right)\right].		
	\end{split}
\end{equation}
We can already pause here for a comment that is going to be extremely important in the following.
At the vacuum where $M_{AB}(x)=\delta_{AB}$, $U$ and $U_0$ generate the same O(6,6) matrix ${\cal H}_{MN}$, from which the 10-dimensional metric, 2-form and dilaton follow via (\ref{10dorigin})--(\ref{10dorigfin}).
However the SO(6) $\times$ SO(6) matrix $R$ is crucial to obtain a twist matrix $U$ 
 that satisfies the consistency constraints required to give a consistent truncation to 4 dimensions.
Only $U$ gives constant $f_{ABC}$ elements via (\ref{cons1}), while $U_0$ produces coordinate dependent structure constants, which are unacceptable for our reduction procedure.
Using the uplift formulae above, the twist matrix generates a background with metric
\begin{equation}
	\begin{split}
	ds^2 = 2 e^{-\phi/2} [&d \theta_1^2 +\cos^2 \theta_1 \, d \theta_2^2  +\sin^2 \theta_1 \,d \theta_3^2 +d \psi_1^2 \\[2mm]
	&\left.+(1+ {\rm{sech}}(2 \psi_1))d \psi_2^2 + (1-{\rm{sech}}(2 \psi_1))  \,d \psi_3^2\right],
	\end{split}
\end{equation}
dilaton
\begin{equation}
	e^\phi = \frac{2}{\sqrt{\cosh(2 \psi_1)}}
\end{equation}
and B-field
\begin{equation}
	B = 4 e^{\phi/2} \left[\sin^2 \theta_1 \, d \theta_2 \wedge d \theta_3 + \frac{\sinh \psi_1}{\cosh^2 (2\psi_1)} \, d \psi_2 \wedge d \psi_3\right]
\end{equation}
It is therefore clear that the starting point of our analysis is a regular differentiable manifold that is the product of a sphere and a hyperboloid.



\section{Going to the boundary of the moduli space} 
\label{sec:going_to_the_boundary_of_the_moduli_space}

Now that we have all the formulas and ingredients to relate the bosonic sector of $N=4$, $d=4$ gauged supergravity to 10-dimensional supergravity and we reviewed the construction of the $S^3 \times {\cal H}^{2,2}$ background found in \cite{Baguet:2015iou,Malek:2017cle}, we proceed to the analysis of the deformations obtained by taking to the boundary any modulus of the corresponding Minkowski$_4$ solution.
First we prove that singular limits of gauged supergravities admitting a generalized Scherk--Schwarz uplift are themselves upliftable.
This guarantees that our procedure gives a regular background of 10-dimensional supergravity. 
Then we give a simple example of a singular limit for the GL$^+(4)$ generalized parallelization of $S^3$, where the limit procedure can be followed explicitly step by step,
In the next section we then apply it to the $S^3 \times {\cal H}^{2,2}$ vacuum describing in detail the new limit vacua.

\subsection{Generalized Scherk--Schwarz uplifts of gauged supergravities} 
\label{sub:generalized_scherk_schwarz_uplifts_of_gauged_supergravities}

The first part of our discussion is a quick review of the general construction of \cite{Inverso:2017lrz} (see also \cite{duBosque:2017dfc} for earlier work) that provides the conditions for the existence of a generalized Scherk--Schwarz uplift of a gauged supergravity theory.

Given a supergravity theory with global symmetries $\mathcal G\times\mathbb R^+$, broken by a gauging defined by an embedding tensor $`X_AB^C`$, one can uplift the theory to a higher-dimensional supergravity by means of an extended field theory (such as DFT and ExFT) based on the same duality group $\mathcal G\times\mathbb R^+$, if certain conditions are met which we now describe.
The internal space will necessarily be a coset space $\widehat{\mathrm{G}}/\widehat{\mathrm H}$ constructed in terms of the centrally extended gauge group $\widehat{\mathrm G}$, defined by formal generators $\widehat X_A$ satisfying
\begin{equation}
	\label{centrally extended gauge algebra}
`X_(AB)^C`\widehat X_C = 0\,,\qquad 
[\widehat X_A,\,\widehat X_B] = -`X_AB^C` \widehat X_C\,,
\end{equation}

We can then associate to these formal generators $\widehat X_A$ a centrally extended embedding tensor $\widehat\Theta_A{}^{\hat a}$ satisfying $`X_(AB)^C`\widehat\Theta_C{}^{\hat a}=0$, where $\hat a$ runs along the $\widehat{\mathrm G}$ coadjoint representation.
For the uplift to exist, the projection $\widehat\Theta_A{}^{\underline{m} }$ onto a set of coset generators $\{\hat t_{\underline{m} }\}$ must satisfy the section condition
\begin{equation}\label{theta section}
`Y^AB_CD`\widehat\Theta_A{}^{\underline{m} }\widehat\Theta_B{}^{\underline{n} } = 0\,
\end{equation}
and the physical internal derivatives of ExFT are then identified as%
\footnote{There is an extra `C-like' condition that needs to be imposed for general extended field theories, but it does not play any role in the proof and it is redundant for double and exceptional field theories as long as we uplift to ten and/or eleven dimensions and the gauging does not involve the higher dimensional trombone symmetry ${\mathbb R}^+_0$.}
\begin{equation}\label{theta gives section}
\partial_M \equiv \widehat\Theta_M{}^m \frac\partial{\partial y^m}\,.
\end{equation}
This choice of section breaks the global symmetry group ${\mathcal G}\times{\mathbb R}^+$ of ExFT  down to the semidirect product of the GL$(d)$ structure group of the internal manifold, the global symmetries ${\mathcal G}_0\times{\mathbb R}^+_0$ of the higher dimensional theory (${\mathbb R}^+_0$ being its trombone symmetry), and shifts of the internal $p$-form potentials, forming a unipotent subgroup we denote by ${\mathcal P}$.%
\footnote{In general the GL(1) centre of the structure group, corresponding to overall rescalings of the internal coordinates, does not belong to ${\mathcal G}$ but is rather a linear combination of some GL$(1)'\subset{\mathcal G}$ and the trombone ${\mathbb R}^+$. See e.g. \cite{Samtleben:2008pe}.}
Using equations:
\begin{equation}
A_M{}^N \widehat\Theta_N{}^m  = \widehat\Theta_M{}^n g_n{}^m\,,\quad
A_M{}^N \in ({\mathrm GL}(d)\times{\mathcal G}_0\times{\mathbb R}^+_0)\ltimes{\mathcal P} \,,\quad
g_n{}^m \in {\mathrm GL}(d)\,.
\end{equation}
In particular, ${\mathrm GL}(d)\ltimes{\mathcal P}$ is the (split) structure group of the extended generalized tangent bundle associated with the choice of section \eqref{theta gives section}.
The generalized Scherk--Schwarz uplift for this gauged supergravity is then encoded into a generalized frame/twist matrix $\hat E(y)_A{}^M\in{\mathcal G}\times{\mathbb R}^+$ satisfying \eqref{gSS condition}.
The latter has the universal form 
\begin{equation}\label{genSS frame}
\hat E(y)_A{} ^M = L(y)^{-1}_A{}^B \mathring e(y)_B{}^N C(y)_N{}^M\,,
\end{equation}
where $L(y)$ is a coset representative\footnote{Despite the notation, we take the coset space as a \textit{left} coset: $L\simeq h L$ with $h \in \widehat{\mathrm H}$.} for $\widehat{\mathrm G}/\widehat{\mathrm H}$ (embedded into ${\mathcal G}\times{\mathbb R}^+$, so that the central extensions are trivially represented), $\mathring e(y)_B{}^N$ is the inverse reference vielbein obtained by projection of the Cartan--Maurer form ${\mathrm d} L L^{-1}$ onto the coset generators $\hat t_{\underline{m}}$, (embedded in the duality group as a GL$(d)$ matrix $\subset{\mathcal G}\times{\mathbb R}^+$, where GL$(d)$ is the structure group of the internal manifold), and $C(y)_N{}^M$ satisfies
\begin{equation}
C(y)_N{}^M \widehat\Theta_M{}^m = \widehat\Theta_N{}^m\,,
\end{equation}
and encodes in particular the information on the $p$-form potentials in the generalized Scherk--Schwarz reduction.
Crucially, a suitable $C(y)_N{}^M$ such that \eqref{genSS frame} solves \eqref{gSS condition} can always be constructed and \eqref{genSS frame} can be proven to define a \emph{global} generalized Leibniz parallelisation of the generalized tangent bundle \cite{Inverso:2017lrz}.


\subsection{Singular limits of generalized Scherk--Schwarz reductions} 
\label{sub:singular_limits_of_generalized_scherk_schwarz_reductions}

Given these conditions, we now want to modify the embedding tensor defining a given consistent background by introducing appropriate rescalings related to the expectation value of certain moduli fields and check under which conditions the result remains a consistent background, also in the limit of an infinite deformation.
We start by introducing the deformed embedding tensor
\begin{equation}
X^\xi_{AB}{}^C \equiv \xi^p \, `G_A^D``G_B^E` `X_DE^F` `G-1_F^C`\,,
\end{equation}
It will be more convenient to reabsorb $\xi^p$ as a trombone component for $G(\xi)$, so that we define
\begin{equation}
{V_\xi} \equiv \xi^p G(\xi) \in {\mathcal G}\times{\mathbb R}^+\,,\qquad 
X^\xi{}_{AB}{}^C = {V_\xi}_A{}^D{V_\xi}_B{}^E `X_DE^F` {V_\xi}^{-1}{}_F{}^C
\end{equation}
As long as $\xi$ is finite, $X^\xi$ sits in the same duality orbit as $X$ and we can uplift it by a twist matrix that is simply $V_\xi \hat E$.
However, when we take the value of $\xi$ to its boundary limit, this twist matrix becomes singular and does not define a generalized Scherk--Schwarz uplift for $X_{\text{contr}}$.

We will now prove that the twist matrix $V_\xi \hat E$ can be rendered non-singular in the limit sending $\xi$ to the boundary of the moduli space by performing a change of coordinates and by making an appropriate gauge choice for the $p$-form potentials encoded within it, combined with a constant, $\xi$ dependent action of the global symmetries ${\mathcal G}_0\times{\mathbb R}^+_0$ of the higher dimensional theory.
To do so, we shall use the universal form \eqref{genSS frame} of the generalized frame.
First of all, let us stress that we consider the limit in $\xi$ to really be a limit along the scalar manifold ${\mathcal G}/{\mathcal H}$, which means that $V_\xi$ can be parameterised in any ${\mathcal H}$ gauge and any two gauge choices must be related by an ${\mathcal H}$ transformation with a well-defined (non oscillating) limit. 
In particular, this means that we can always pick $V_\xi$ to belong to the (block) triangular subgroup of ${\mathcal G}\times{\mathbb R}^+$ that preserves the choice of section:
\begin{equation}\label{V in gen struct group}
V_\xi{}_A{}^B \widehat\Theta_B{}^{\underline{m} } = \widehat\Theta_A{}^{\underline{n} } g_\xi{}_{\underline{m} }{}^{\underline{n} }\,,
\end{equation}
Which means that $V_\xi \in ({\mathrm{GL}}(d)\times{\mathcal G}_0\times{\mathbb R}^+_0)\ltimes{\mathcal P}$.
This gauge choice simplifies the following proof, but we will see that we can drop it at the very end.

We now decompose the embedding tensor in terms of coset and H generators (denoted $t_{\underline{m} }$ and $t_i$ respectively):
\begin{equation}
`X_AB^C` = `\Theta_A^{\underline{m} }``t_{\underline{m} }B^C` + `\Theta_A^i``t_iB^C`\,.
\end{equation}
These generators are embedded in the duality algebra and hence the central extension is represented trivially.
If we now look at the deformed embedding tensor, we can write
\begin{equation}\label{expand X lambda}
`X\xi_AB^C` = 
`\Theta_A^{\underline{m} }` g_\xi{}_{\underline{m} }{}^{\underline{n} } (V_\xi t_{\underline{n} }V^{-1}_\xi)_B{}^C 
+ `V_\xi\,_A^D` `\Theta_D^i`(V_\xi t_i V^{-1}_\xi)_B{}^C  \,.
\end{equation}
The first term in this expansion is separately finite in the $\xi$ limit, at least for a special choice of coset generators.
Indeed, introducing the orthogonal projector $\Pi$ onto the image of $\widehat\Theta_A{}^{\underline{m} }$ and its complement $\overline\Pi$ (so that $\Pi_A{}^B\widehat\Theta_B{}^{\underline{m} }=\widehat\Theta_A{}^{\underline{m} }$, $\overline\Pi_A{}^B\widehat\Theta_B{}^{\underline{m} }=0$), we notice that they are invariant under the action of $V_\xi$ satisfying \eqref{V in gen struct group}.
For any $\xi$, there is a choice of coset generators\footnote{The factorised $g_\xi$ is there so that $\hat t^\xi_{\underline{m} }$ satisfy the same commutation relations for any finite $\xi$.} $g_\xi{}_{\underline{m} }{}^{\underline{n} }\hat t^{\xi}_{\underline{n} }$ such that $\Pi_A{}^B \widehat X_B = \widehat\Theta_A{}^{\underline{m} }g_\xi{}_{\underline{m} }{}^{\underline{n} } \hat t^{\xi}_{\underline{n} }$ and in particular this will match with the first term \eqref{expand X lambda} when embedded into ${\mathcal G}\times{\mathbb R}^+$, with $t_{\underline{m} }{}_A{}^B = \hat t^{\xi=1}_{\underline{m} }{}_A{}^B$.
Being this the contraction of two objects both finite under the $\xi$ limit, the first entry of \eqref{expand X lambda} is finite.

This choice of coset generators allows us to construct a twist matrix for $X^\xi$ that is finite in the limit and  matches the original twist matrix \eqref{genSS frame} for $\xi=1$.
We begin by making a specific choice for the coset representative of ${\mathrm G}/{\mathrm H}$:
\begin{equation}\label{choice of coset rep 1}
L_\xi(y) \equiv \exp(y^m g_{\xi\,m}{}^n \hat t_n^\xi )
\quad
\overset{\xi\to1}{\longrightarrow}
\quad 
[L(y)]=\exp\left(y^m\hat t_m\right)
\end{equation}
Notice that on the right hand side we have made a special choice of coordinates and H gauge for the original twist matrix.
Embedding into ${\mathcal G}\times{\mathbb R}^+$, we can rewrite this definition as
\begin{equation}\label{choice of coset rep 2}
[L_\xi(y)]_A{}^B = [ V_\xi L(y\cdot g_\xi) V^{-1}_\xi]_A{}^B\,.
\end{equation}
Notice how $g_\xi$, defined in \eqref{V in gen struct group} as the GL$(d)$ component of $V_\xi$, appears as a change of coordinates.
Computing a reference vielbein by projecting ${\mathrm d} L_\xi L_\xi^{-1}$ onto the coset generators $(g_\xi)_{\underline{m} }{}^{\underline{n} } \hat t_{\underline{n} }^\xi$, we find
\begin{equation}
\mathring e_\xi(y) = g_\xi \mathring e(y\cdot g_\xi) g_\xi^{-1}\,,
\end{equation}
where the right action of $g_\xi$ is due to the redefinition of the coset generators, while its other appearances can be interpreted as a change of coordinates $y \to y\cdot g_\xi$.
At this point we construct a tentative twist matrix following \cite{Inverso:2017lrz} as
\begin{equation}\label{finite tentative frame}
E^{\xi,\,tent.}(y){}_A{}^M \equiv
 [L_\xi(y)^{-1}]_A{}^B [\mathring e_\xi(y)]_B{}^M\,,
\end{equation}
which is finite in the $\xi$ limit and reduces to $\hat E(y)_A{}^N C^{-1}(y)_N{}^M$ (defined in \eqref{genSS frame}) if we set $\xi = 1$.
Notice that we are keeping our choice of section fixed to $\widehat\Theta_M{}^m$.
Expanding $L_\xi$ and $\mathring e_\xi$, we arrive at
\begin{equation}
	\begin{split}
		E^{\xi,\,tent.}(y){}_A{}^M &=
		[V_\xi L^{-1}(y\cdot g_\xi)]_A{}^B [\mathring e(y\cdot 	g_\xi)g_\xi^{-1}]_B{}^N R_\xi(y)_N{}^M\\[1ex]
		&= V_\xi{}_A{}^B [\hat E\, C^{-1}](y\cdot g_\xi)_B{}^N [g^{-1}_\xi]_N{}^P R_\xi(y)_P{}^M
	\end{split}
\end{equation}
with $R_\xi(y) \in ({\mathcal G}_0\times{\mathbb R}^+_0)\ltimes{\mathcal P}$ being equal to $V_\xi^{-1}g_\xi$ conjugated with $\mathring e(y\cdot g_\xi)g_\xi^{-1}$.
In general, $E^{\xi,\,tent.}(y)$ does not satisfy the generalized Scherk--Schwarz conditions, neither before nor after the $\xi$ limit. 
However, the general construction in \cite{Inverso:2017lrz} tells us that there always exists some matrix $C_\xi(y)\in ({\mathcal G}_0\times{\mathbb R}^+_0)\ltimes{\mathcal P}$ that completes $E^{\xi,\,tent.}(y)$ to the correct twist matrix, just like in \eqref{genSS frame}.
This can be computed by integrating the generalized flux defined as the difference between $X^\xi$ and the generalized torsion of $E^{\xi,\,tent.}(y)$.
We will not need any explicit expression.
We only stress that since $E^{\xi,\,tent.}(y)$ is finite in the limit, $C_\xi(y)$ is as well, at least for some choice of gauge.
We therefore conclude that there exists a twist matrix $E^{\xi,\,tent.}(y) C_\xi(y)$ that correctly reproduces $X^\xi$.
However, at finite $\xi$, for our choice of section and of coset space we have already found such a twist matrix: it is just $V_\xi \hat E(y)$!
We conclude that the two can only differ by a finite generalized diffeomorphism and by a global symmetry transformation in the higher dimensional theory, as anticipated.
The generalized diffeomorphism is composed of the change of coordinates $y\to y\cdot g_\xi$ and of $p$-form gauge transformations  encoded into the ${\mathcal P}$ component of
\begin{equation}\label{pure gauge}
\Lambda(y) \equiv  g_\xi C^{-1}(y\cdot g_\xi) g^{-1}_\xi \, R_\xi(y) \,  C_\xi(y) 
\,,
\end{equation}
which is therefore pure gauge, i.e. has vanishing generalized torsion.
The global symmetry transformation corresponds to the ${\mathcal G}_0\times{\mathbb R}^+_0$ component of $\Lambda(y)$, which is indeed constant.
The resulting twist matrix 
\begin{equation}\label{final finite twist matrix}
\hat E^\xi_{\text{finite}}(y)_A{}^M \equiv [V_\xi \hat E(y\cdot g_\xi) g^{-1}_\xi \Lambda(y)]_A{}^M
\end{equation}
therefore reproduces $X^\xi$ and is finite in the limit, where it reproduces the contracted embedding tensor.
It differs from the natural deformed twist matrix $V_\xi \hat E(y)$ only by a change of coordinates, $p$-form gauge transformations and constant, $\xi$ dependent global symmetry transformations of the higher dimensional theory.

We stress that the proof above is constructive: given the choice of coordinates specified in \eqref{choice of coset rep 1}, \eqref{choice of coset rep 2}, the change of coordinates is just the GL$(d)$ component of $V_\xi$, while \eqref{pure gauge} can be computed by integration of the generalized flux described in \cite{Inverso:2017lrz}.
In practice, however, it can be more convenient to reconstruct such coordinate and gauge transformations on a case-by-case basis, knowing that they must exist.

A final comment is in order on the rescaling of the gauge coupling associated with the ${\mathbb R}^+$ trombone component $\xi^p$ of $V_\xi = \xi^p G(\xi)$.
Whenever $p\neq0$, $\Lambda(y)$ contains necessarily a constant component in ${\mathbb R}^+_0$ that descends from $V_\xi^{-1}g_\xi$ in the definition of $R_\xi$. 
We remind the reader that ${\mathbb R}^+_0$ is the trombone symmetry of the higher dimensional theory and it is a linear combination of ${\mathbb R}^+$ (the lower-dimensional trombone) and of some ${\mathrm GL}(1)'\subset{\mathcal G}$.
The linear combination is such that the internal derivatives are invariant just like in \eqref{V in gen struct group}.
We can therefore decompose $V_\xi$ on the left of \eqref{final finite twist matrix} into $\xi^p G(\xi)$, and bring the trombone component to the right, where it will combine with the ${\mathbb R}^+_0$ component of $\Lambda(y)$ to give rise to a constant ${\mathrm GL}(1)'$ trasformation.
We thus write
\begin{equation}\label{final finite twist matrix 2}
\hat E^\xi_{\text{finite}}(y)_A{}^M = [G(\xi) \hat E(y\cdot g_\xi) g^{-1}_\xi \Lambda'(y) T_\xi]_A{}^M
\end{equation}
where $G(\xi)\in{\mathcal G}$ represents a curve in scalar field space, $\Lambda(y)'$ is $\Lambda(y)$ stripped of its ${\mathbb R}^+_0$ component, and $T_\xi\in{\mathrm GL}(1)'\subset{\mathcal G}$ acts on internal derivatives (on section) as
\begin{equation}
(T_\xi)_M{}^N \partial_N = \xi^p \partial_M\,,
\end{equation}
therefore reproducing the rescaling of the gauge coupling in the deformed embedding tensor.

Finally, we now notice that we do not need $G(\xi)$ on the left hand side of \eqref{final finite twist matrix 2} to be in the ${\mathcal H}$ gauge defined by \eqref{V in gen struct group}.
This is so because we are taking a singular limit along the scalar field space ${\mathcal G}/{\mathcal H}$, and thus the choice of ${\mathcal H}$ gauge must not matter in the limit.
This implies that any other choice of gauge is related to \eqref{V in gen struct group} by an ${\mathcal H}$ transformation that has a well-definite (not oscillating) $\xi$ limit.

Summarizing, we find that contractions of a gauging obtained through singular limits in scalar field space always admit an uplift if the original gauging does.
The twist matrix for the contracted model is obtained from the following steps:
\begin{enumerate}
\item multiply from the left by the modulus transformation $G(\xi)$ parameterizing a path to the boundary of ${\mathcal G}/{\mathcal H}$;
\item perform a $\xi$ dependent change of coordinates;
\item implement a $\xi$ dependent gauge transformations for the $p$-forms;
\item employ a constant, $\xi$ dependent action of the higher dimensional symmetry group ${\mathcal G}_0\times{\mathbb R}^+_0$;
\item perform a constant, $\xi$ dependent GL(1)$'$ transformation to reproduce the rescaling of the gauge coupling constant. 
\end{enumerate}

The above result is fully general. 
Specialising now to O(6,6) DFT, the constant $\xi^p$ scaling should be always regarded as a 4d trombone transformation in order to reproduce the correct (finite) limits for all fields. 
Because we are regarding the twist matrix to be O(6,6) valued, it is simpler to rely on \eqref{final finite twist matrix 2} and regard the $\mathrm{GL}(1)'$ scaling as being generated by ${\mathbbm 1}_6\oplus-{\mathbbm 1}_6$.
In fact, $\rho(y)$ is also affected by the $\mathrm{GL}(1)'$ scaling as well as by the change of coordinates, but instead of looking at its transformation explicitly, we can define it by integration of \eqref{cons2} and it will be automatically finite in the limit.
We also find that step 4 is unnecessary in our examples, as the only $\mathcal G_0$ transformation that would be needed is a shift of the ten-dimensional dilaton which is however automatically taken into account by the integration of \eqref{cons2}.
Furthermore, step 5 boils down to the SO(6,6) transformation $\left(\begin{smallmatrix}\xi^p{\mathbbm 1}_6&\\&\xi^{-p}{\mathbbm 1}_6\end{smallmatrix}\right)$ acting on $U^{-1}_A{}^M$ from the right.


\subsection{A simple example: $S^3$ generalized Scherk--Schwarz reduction and CSO(2,0,2) limit} 
\label{sub:simple_examples}

In order to make things concrete, we now apply the general limit procedure described in the previous section to the simple example of the O(3,3) generalized Scherk--Schwarz reduction on a three-sphere and its singular limit to a twist matrix giving a CSO(2,0,2) gauging.

The first step is the construction of the generalized parallelization of the three-sphere, based on an SL(4) $\simeq$ SO(3,3) twist matrix \cite{Lee:2014mla}.
$S^3$ can be defined by embedding coordinates in ${\mathbb R}^4$ satisfying $\sum_{a=1}^4 (Y^a)^2=1$:
\begin{equation}\label{S3 embcoords}
	\begin{split}
		Y^1 &= \cos\theta_1 \cos\theta_2\,,\qquad	Y^2 = \cos\theta_1 \sin\theta_2\,,\\
		Y^3 &= \sin\theta_1 \cos\theta_3\,,\qquad	Y^4 = \sin\theta_1 \sin\theta_3\,.
	\end{split}
\end{equation}
The reference metric on $S^3$ is (we use $\theta^i = (\theta_1,\,\theta_2,\,\theta_3)$)
\begin{equation}\label{S3 ref metric}
\mathring g_{ij} =
\partial_i Y^a \partial_j Y^a\,,\qquad
{\mathrm d} \mathring s^2 =
{\mathrm d}\theta_1^2+ \cos\theta_1{\mathrm d}\theta_2^2 +\sin\theta_1{\mathrm d}\theta_3^2\,.
\end{equation}
Coordinate ranges are $0<\theta_1<\pi/2$, $0\le\theta_{2,3}<2\pi$.
We also have $\mathring g^{1/4} = \sqrt{\cos\theta_1\sin\theta_1}$.
To construct the twist matrix we also need $B^i\equiv\frac12 \varepsilon^{ijk}B_{jk}$.
It is crucial to make a gauge choice that is non-singular in the limits we want to take. 
For the moment let us take
\begin{equation}\label{S3 Bmu}
B^i = \left(\tfrac12 \sin^2\!\theta_1\,,\ 0\,,\ 0\right)\,.
\end{equation}
We will need to tune this gauge choice depending on the singular limit we want to take, in order to avoid diverging pure-gauge terms.

The twist matrix in the $\mathbf4$ of SL(4) can be written as ($m=i,4$)
\begin{equation}\label{S3 fundFrame}
U^{-1}{}_a{}^m = 
\begin{pmatrix}
\mathring g^{1/4}\mathring g^{ij}\partial_j Y_a 
+ \mathring g^{-1/4} B^i\ , &
\mathring g^{-1/4} Y_a
\end{pmatrix}\,.
\end{equation}
where $Y_a = \eta_{ab}Y^b$ and $\eta_{ab}=\delta_{ab}$.
We can then define the twist matrix in the $\mathbf 6$ of SL(4) by tensoring:
\begin{equation}\label{S3 Frame v0}
U^{-1}{}_{ab}{}^{mn} \equiv U^{-1}{}_{[a}{}^{[m}U^{-1}{}_{b]}{}^{n]}\,.
\end{equation}
In this representation the SL(4)$\simeq$SO(3,3) invariant is $\eta_{ab,cd}=\varepsilon_{abcd}$.
We can reorder the indices to switch to more common conventions for SO(3,3).
Define $E_A{}^M$ with $A,\,M=1,...,6$ in three-by-three blocks
\begin{equation}\label{S3 Frame O33}
U^{-1}{}_A{}^M \equiv
\begin{pmatrix}
 U^{-1}{}_{a4}^{m4} \ & 
 \tfrac12 U^{-1}{}_{a4}^{np}\varepsilon_{npm4} \\
 \tfrac12 \varepsilon^{a4bc}  U^{-1}{}_{bc}^{m4}\ &
  \tfrac14 \varepsilon^{a4bc} U^{-1}{}_{bc}^{np} \varepsilon_{npm4}
\end{pmatrix}\,.
\end{equation}
The associated SO(3,3) invariant is now $\eta_{AB}=\eta_{MN}=\left(\begin{smallmatrix}	&{\mathbbm 1}_3\\{\mathbbm 1}_3 &\end{smallmatrix}\right)$.
The twist matrix $U^{-1}{}_A{}^M$ satisfies \eqref{cons1} with nonvanishing embedding tensor components
\begin{equation}\label{S3 emb tens}
`f_AB^C`\sim\varepsilon_{ABC}\quad\text{for}\quad A,B,C=1,2,3\,,\ \text{or}\  A,B,C=4,5,6\,.
\end{equation}

As a warm-up, let us investigate the limiting procedure to CSO(2,0,2).
We start by acting with the SO(3,3) transformation
\begin{equation}\label{S3 to CSO202 scaling}
`\Lambda_A^B` = \mathrm{diag}\left(1,\,1,\,\tfrac1\xi,\,1,\,1,\,\xi\right)
\end{equation}
which, in the ${\mathbf 4}$ of SL(4), is $\mathrm{diag}(\sqrt\xi,\,\sqrt\xi,\,1/\sqrt\xi,\,1/\sqrt\xi)$.
The singular limit on the embedding tensor is obtained by the action of $\Lambda$ combined with an overall rescaling
\begin{equation}\label{SO4 to CSO202 emb tens limit}
\lim_{\xi\to+\infty} f^{(\xi)}_{AB}{}^C \equiv \lim_{\xi\to+\infty} \frac1\xi`\Lambda_A^D``\Lambda_B^E``f_DE^F``\Lambda-1_F^C` \,.
\end{equation}
To implement this limit in terms of the twist matrix, we define
\begin{equation}\label{S3 to CSO202 defo E}
U^{-1}_{(\xi)}{}_A{}^M = \Lambda_A{}^B U^{-1}_B{}^P `T_P^M`
\end{equation}
where $`T_M^N` = \mathrm{diag}(1/\xi,1/\xi,1/\xi,\xi,\xi,\xi)$ is the $\mathrm{GL}(1)'$ transformation discussed in the previous section, implementing the overall gauge coupling scaling.
Notice that $\det U = \det U_{(\xi)} = 1$.
The string frame metric ($i,\,j$ are the first three entries of the indices $M,\,N$) is then easily derived from $U_{(\xi)}$ by
\begin{equation}\label{SO4 to CSO202 defo metric}
	\begin{split}
		g^{(\xi)}_{ij} &= ({\mathcal H}_{(\xi)}^{ij})^{-1} = (U^{-1}_{(\xi)}{}_A{}^i U^{-1}_{(\xi)}{}_A{}^j)^{-1}\,,\\
		{\mathrm d} s^2 &= \xi^2 \left[ \,
		{\mathrm d}\theta_1^2 \, + (\xi^2-1+\cos^{-2}\!\theta_1)^{-1} {\mathrm d}\theta_2^2
		\, + \left(\tfrac{1}{\xi^2}-1+\sin^{-2}\!\theta_1\right)^{-1} {\mathrm d}\theta_3^2 \,
		\right] 
	\end{split}
\end{equation}
and the $B$-field is identified with
\begin{equation}\label{SO4 to CSO202 defo B}
B_{(2)} = -\xi^2 (\xi^2-1+\cos^{-2}\!\theta_1)^{-1} \tan^2\theta_1\ {\mathrm d}\theta_2\wedge{\mathrm d}\theta_3\,.
\end{equation}
The overall $\xi^2$ factors are due to the $T$ rescaling in both the metric and $B$ field.
Reading these expressions one notes immediately that the $\xi \to +\infty$ limit is singular, unless one combines it with simultaneous change of coordinates $\theta^1\to\theta^1/\xi$, whose associated Jacobian is embedded into SO(3,3) as follows:
\begin{equation}\label{SO4 to CSO202 jacobian}
`j_\mu^\nu` = \frac{\partial \theta'^\nu}{\partial\theta^\mu} = \mathrm{diag}(1/\xi,1,1)\,,\qquad
`J_M^N` = 
\begin{pmatrix}
`j_\mu^\nu` & \\ & `j-1^\mu_\nu`
\end{pmatrix}\,.
\end{equation}
Note that $J$ and $T$ commute.
Under the change of coordinates $U^{-1}_{(\xi)}$ therefore transforms as
\begin{equation}\label{rdukxy}
U^{-1}_{(\xi)}(\theta^1,\,\theta^2,\,\theta^3) \to U^{-1}_{(\xi)}(\theta^1/\xi,\,\theta^2,\,\theta^3)\, J^{-1}\,.
\end{equation}
Note also that the change of coordinates has no effect on the associated torsion $f^{(\xi)}_{AB}{}^C$, but guarantees that the twist matrix remains non-singular in the limit.

The  $\xi\to +\infty$ limit now gives a regular result, with a vanishing $B$-field and a \string frame metric 
\begin{equation}\label{sxdcfghv}
({\mathrm d} s^{(\infty)})^2 = {\mathrm d}\theta_1^2 +{\mathrm d}\theta_2^2 +\theta_1^2\,{\mathrm d}\theta_3^2\,,
\end{equation}
which is a flat metric in cylindrical coordinates.

Let us also look at the explicit form of the twist matrix before and after the limit:
\begin{align}\label{hsdf}
&E_{\scriptscriptstyle S^3}^{(\xi)}(\theta^1/\xi,\,\theta^2,\,\theta^3)\, J^{-1} =\\[1ex]\nonumber
&=\left(\begin{smallmatrix}
-\cos\theta_2\sin\theta_3 
& -\frac1\xi\sin\theta_2\sin\theta_3\tan\frac{\theta_1}{\xi}
& -\frac1\xi\cos\theta_2\cos\theta_3\cot\frac{\theta_1}{\xi}
& -\sin\theta_2\cos\theta_3
& 0
& \xi\sin\theta_2\sin\theta_3\tan\frac{\theta_1}{\xi}\\
-\sin\theta_2\sin\theta_3 
& -\frac1\xi\cos\theta_2\sin\theta_3\tan\frac{\theta_1}{\xi}
& -\frac1\xi\sin\theta_2\cos\theta_3\cot\frac{\theta_1}{\xi}
& \cos\theta_2\cos\theta_3
& 0
& -\xi\cos\theta_2\sin\theta_3\tan\frac{\theta_1}{\xi}\\
0&0&-\frac1{\xi^2}&0&-1&0\\
-\sin\theta_2\cos\theta_3 
& \frac1\xi\cos\theta_2\cos\theta_3\tan\frac{\theta_1}{\xi}
& \frac1\xi\sin\theta_2\sin\theta_3\cot\frac{\theta_1}{\xi}
& -\cos\theta_2\sin\theta_3
& 0
& -\xi\cos\theta_2\cos\theta_3\tan\frac{\theta_1}{\xi}\\
-\cos\theta_2\cos\theta_3 
& \frac1\xi\cos\theta_2\sin\theta_3\tan\frac{\theta_1}{\xi}
& -\frac1\xi\cos\theta_2\sin\theta_3\cot\frac{\theta_1}{\xi}
& -\sin\theta_2\sin\theta_3
& 0
& -\xi\sin\theta_2\cos\theta_3\tan\frac{\theta_1}{\xi}\\
0&-1&0&0&0&0
\end{smallmatrix}\right)\\[2ex]\nonumber
&\overset{\scriptscriptstyle\xi\to+\infty}{\longrightarrow}
\left(\begin{smallmatrix}
-\cos\theta_2\sin\theta_3 
& 0
& -\frac1{\theta_1}\cos\theta_2\cos\theta_3
& -\sin\theta_2\cos\theta_3
& 0
& \theta_1 \sin\theta_2\sin\theta_3\\
-\sin\theta_2\sin\theta_3 
& 0
& -\frac1{\theta_1}\sin\theta_2\cos\theta_3
& \cos\theta_2\cos\theta_3
& 0
& -{\theta_1}\cos\theta_2\sin\theta_3\\
0&0&0&0&-1&0\\
-\sin\theta_2\cos\theta_3 
& 0
& \frac1{\theta_1}\sin\theta_2\sin\theta_3
& -\cos\theta_2\sin\theta_3
& 0
& -{\theta_1}\cos\theta_2\cos\theta_3\\
-\cos\theta_2\cos\theta_3 
& 0
& -\frac1{\theta_1}\cos\theta_2\sin\theta_3
& -\sin\theta_2\sin\theta_3
& 0
& -{\theta_1}\sin\theta_2\cos\theta_3\\
0&-1&0&0&0&0
\end{smallmatrix}\right)\,.
\end{align}
The coordinate change crucially modified the twist matrix so that the result after the limit is regular and non-trivial.
We shall see for the full examples in the next section that upon switching to Cartesian coordinates, such a twist matrix is orthogonal and can be interpreted in terms of an asymmetric orbifold.



\section{New background limits} 
\label{sec:new_background_limits}

We now apply the procedure described in the previous section to the $S^3 \times {\cal H}^{2,2}$ compactification described in section \ref{sec:ten_dimensional_origin} performing limits to the boundary of the moduli space for all the moduli of the corresponding 4-dimensional gauged supergravity.
In general, assigning a finite value to a modulus $e_i$ when taking the singular limit along the associated $x_1$ can yield inequivalent gaugings \cite{Catino:2013ppa}.
In studying the ten-dimensional solutions arising from such setups we have found that they are qualitatively analogous to the cases where one leaves $e_i=1$ in the $x_i$ limit.
We therefore choose to present only this smaller class of solutions.
We will focus on the twist matrix $U$, which is at the basis of the 10-dimensional uplift. 
We especially remind that the background 10-dimensional internal metric and 2-form $B$ are specified by 
\begin{equation}
	{\cal H}_{MN}(y) = U_M{}^A(y) \delta_{AB} U_N{}^B(y)
\end{equation}
and their deformations follow the appearance of non-trivial vevs to $M_{AB}(x)$, which can in turn be defined in terms of SO(6,6)/(SO(6) $\times$ SO(6)) scalar $\sigma$-model representatives $L$ as $M = L L^T$.
This means that we can introduce the moduli-dependent twist $U(x,y) = U(y) L(x)$ and study its limits.
For the [SU(1,1)/U(1)]$^2$ moduli space described in section \ref{sec:the_moduli_space_of_minkowski_vacua_in_n_8_supergravity}, the representative is
\begin{equation}
	L = \exp\left[2 \lambda_2 \, \log e_2\right] \exp\left[2 \lambda_3 \log e_3\right] \exp\left[\sqrt2\, \ell_2\, \log x_2\right] \exp\left[\sqrt2\, \ell_3\, \log x_3\right],
\end{equation}
where we fixed the coefficients such that the boundary of the moduli space is at zero and the fields $e_{2,3}$, $x_{2,3}$ appear with integer powers in the expressions for $U(x,y)$, $\Theta(x)$ etc.

We can summarize the results that we are going to detail in the following in terms of a general structure for the twist matrix and of two classes of metric, B-field and dilaton solutions.
The twist matrix can always be represented as the product of two matrices:
\begin{equation}\label{UR}
	U = U_0 R.
\end{equation}
$R$ is a SO(6) $\times$ SO(6) rotation matrix, which has no effect on the local background geometry, but fixes the non-trivial patching conditions that define the global space.
$U_0$ is the matrix that contains the information on the local background. 
In our examples it is always lower-triangular
\begin{equation}\label{U0}
	U_0 = \left(\begin{array}{cc}
	D &  {\mathbb 0}_6 \\[2mm]
	A & D^{-1} 
	\end{array}\right),
\end{equation}
where $A$ is non-zero only when there is a non-trivial B-field.

The local geometries we obtain are of two types: flat space (with non-trivial patching conditions coming from asymmetric orbifolds) or Q-flux geometries.
While this might not be always immediately clear after the limit, a simple change of coordinates makes this structure explicit.

\subsection{Contractions along $\ell_i$} 
\label{sub:contractions_along_ell_i}

The first series of limits we consider are the limits along the $x_2$, $x_3$ field directions.
We first consider their limits to the boundary of the moduli space separately and then combined.
By doing this, we will produce the uplift of the gaugings listed in the first line of table \ref{tab:contractions}.

\subsubsection{$x_3 \to 0$ limit}

Following the procedure outlined in section \ref{sec:going_to_the_boundary_of_the_moduli_space}, the first regular limit can be obtained by sending $x_3 \to 0$, while rescaling the coordinates 
\begin{equation}
	y^m \to \left\{x_3 \, \theta_1, \theta_2, \theta_3, x_3\, \psi_1, \psi_2, \psi_3\right\}
\end{equation}
and the coupling constant 
\begin{equation}
	g \to \frac{g}{\sqrt2\,x_3}.
\end{equation}
The resulting twist matrix is of the form (\ref{UR})-(\ref{U0}), where $A=0$, 
\begin{equation}
	\begin{split}
	R =& \exp\left[-2 \theta_2 \,C^{13}+2\left(\psi_2+\frac{\pi}{2}\right)\,C^{46}\right]\,\exp\left[2 \left(\theta_3+\frac{\pi}{2}\right) A^{13}+2\left(\psi_3+\frac{\pi}{2}\right)A^{46}\right]  \\[2mm]
	&\, \cdot \exp\left[\pi\left(A^{23}+A^{56}\right)\right].		
	\end{split}
\end{equation}
and
\begin{equation}\label{Dorig}
	D  = {\rm diag}\left\{1,1,\theta_1, 1,1,\psi_1\right\}.
\end{equation}

The background local geometry produced by this solution is that of a flat metric in cylindrical coordinates
\begin{equation}\label{metrx3}
	ds^2 = d \theta_1^2 + d \theta_2^2  + \theta_1^2 \,d \theta_3^2 +d \psi_1^2 + d \psi_2^2 + \psi_1^2 \,d \psi_3^2,
\end{equation}
with constant dilaton $e^\phi = 1$, and vanishing 2-form $B=0$.
Still, the global patching conditions imply that the global solution produces a non-trivial $N=8$ effective theory with supersymmetry-breaking vacua and mass terms generated by a [U(1) $\ltimes T^4]^2$ gauging:
\begin{equation}
	[T_0, T_I] = M_I{}^J T_J, \quad [T_0^\prime, T_I] = N_I{}^J T_J, \quad [T_I, T_J] = 0, \quad [T_0,T_0^\prime] = 0,
\end{equation}
where
\begin{equation}
	M = \frac{{\mathbb 1}+\sigma_3}{2}\otimes {\mathbb 1} \otimes \, i \sigma_2, \quad  	
	N = \frac{{\mathbb 1}-\sigma_3}{2}\otimes {\mathbb 1} \otimes \, i \sigma_2.
\end{equation}
The non-trivial patching conditions become clear once we introduce the coordinates
\begin{equation}\label{coordchange}
		y_1 = \theta_1 \sin \theta_3, \qquad y_2 = \theta_1 \cos \theta_3, \qquad
		y_3 = \psi_1 \sin \psi_3, \qquad y_4 = \psi_1 \cos \psi_3.
\end{equation}
In these coordinates the metric is 
\begin{equation}
	ds^2 = d y_1^2 + dy_2^2 + dy_3^4 + dy_4^4+ d \theta_2^2 + d \psi_2^2,
\end{equation}
and the twist matrix (in the basis $\{y_1,y_2,\theta_2,y_3,y_4,\psi_2\}$) becomes
\begin{equation}
	\label{x3 to 0 cartesian twist matrix}
	U = R = \exp\left[2\theta_2 \, C^{12}-2\left(\psi_2-\frac{\pi}{2}\right)\,C^{45}\right].
\end{equation}
It is now clear that all coordinates can be taken compact and that $U$ is globally well-defined only if there is an asymmetric orbifold acting on the $x_i$ coordinates and their duals, whenever we perform a rotation in $\theta_2$ or $\psi_2$.
Introducing $z = y_1 + i \, y_2$, $w = y_3 + i \,y_4$:
\begin{align}
	\label{orbaction}
 	\left\{ \begin{array}{l}
	\theta_2 \sim \theta_2 + \alpha, \\[2mm]
 	z_L \sim e^{-i \alpha}z_L,\\[2mm]
	z_R \sim e^{i \alpha}z_R,
 \end{array}	\right.
 \qquad
	\left\{ \begin{array}{l}
	\psi_2 \sim \psi_2 + \beta, \\[2mm]
	w_L \sim i\, e^{-i \beta}w_L,\\[2mm]
	w_R \sim -i\, e^{i \beta}w_R.
 \end{array}\right.
\end{align}
We have here an explicit realization of the connection between gauged supergravities and asymmetric orbifold suggested in \cite{Condeescu:2013yma}.
The vacuum generated by the orbifold (\ref{orbaction}) allows for a consistent truncation of the string spectrum such that it describes a spontaneously broken phase of a maximally supersymmetric theory.
This is not guaranteed for any orbifold, nor it is clear that one is always allowed to obtain a consistent truncation to any gauged supergravity theory, especially when supergravity is fully broken like in this case.
In fact we do not have at this stage enough maximal gauged supergravities with Minkowski vacua that could correspond to the string theory vacuum on an (asymmetric) orbifold for any consistent choice of the orbifold action.

\subsubsection{$x_3 \to +\infty$ limit}

The limit to the opposite boundary point in moduli space, namely $x_3 \to +\infty$, gives the same effective theory, with the same gauge group, though involving different combinations of the SO(6,6) generators.
It is however interesting that the resulting background is different from the one just presented.
In order to obtain a regular geometry, one has to send $x_3 \to +\infty$ while rescaling the coordinates 
\begin{equation}
	y^m \to \left\{\theta_1/x_3, \theta_2/x_3^2, \theta_3, \psi_1/x_3, \psi_2/x_3^2, \psi_3\right\},
\end{equation}
and the coupling constant 
\begin{equation}
	g \to g\,\frac{x_3}{\sqrt2}.
\end{equation}
The twist matrix now has the same $D$ as above, 
\begin{equation}
	R = \exp\left(\pi\,C^{46}\right)\,\exp\left[2 \left(\theta_3+\frac{\pi}{2}\right) A^{13}+2\left(\psi_3+\frac{\pi}{2}\right)A^{46}\right] \exp\left[\pi\left(A^{23}+A^{56}\right)\right],
\end{equation}
and 
\begin{equation}
	A = \left(\begin{array}{cccccc}
	0  &  &  &  &  & \\
	&& \theta_1 &&& \\
	&-1 &&&& \\
	&&&0&& \\
	&&&&& \psi_1 \\
	&&&&-1 &
	\end{array}\right),
\end{equation}
so that the resulting local geometry 
\begin{equation}\label{metrx4}
	ds^2 = e^{-\phi/2}\,\left[d \theta_1^2 + \frac{1}{1+ \theta_1^2}(d \theta_2^2  + \theta_1^2 \,d \theta_3^2) +d \psi_1^2 +\frac{1}{1+\psi_1^2} (d \psi_2^2 + \psi_1^2 \,d \psi_3^2)\right]
\end{equation}
has non-zero curvature and a non-trivial dilaton 
\begin{equation}
	e^\phi = [(1+\theta_1^2)(1+\psi_1^2)]^{-1/2}
\end{equation}
and B-field:
\begin{equation}
	B = 2 \frac{\theta_1^2}{1+ \theta_1^2} d \theta_2 \wedge d \theta_3 + 2 \frac{\psi_1^2}{1+ \psi_1^2}d \psi_2 \wedge d \psi_3 .
\end{equation}
Also in this case the global patching conditions become explicit if we perform the (\ref{coordchange}) coordinate change.
After this coordinate change we see that the background is the sum of two copies of Q-flux backgrounds, one in the $(y_1,y_2,\theta_2)$ coordinates and one in the $(y_3,y_4,\psi_2)$ coordinates.
The metric is
\begin{equation}
	\begin{split}
	ds^2 &=e^{-\phi/2}\,\left[ \frac{1}{1+y_1^2+y_2^2} \left(d y_1^2 + dy_2^2 + d \theta_2^2 + (y_1 dy_2 + y_2 dy_1)^2\right) \right. \\[2mm]
	&\left. +\frac{1}{1+y_3^2+y_4^2}\left( dy_3^4 + dy_4^4 + d \psi_2^2 + (y_3 dy_4 + y_4 dy_3)^2\right)\right],
	\end{split}
\end{equation}
with B-field
\begin{equation}
	B = \frac{2}{1+y_1^2+y_2^2}(y_1 dy_2 - y_2 dy_1)\wedge d \theta_2+ \frac{2}{1+y_3^2+y_4^2}(y_3 dy_4 - y_4 dy_3)\wedge d \psi_2
\end{equation}
and dilaton
\begin{equation}
	e^\phi = (1+y_1^2+y_2^2)^{-1/2}(1+y_3^2+y_4^2)^{-1/2}.
\end{equation}
The twist matrix in these coordinates becomes
\begin{equation}
	\begin{small}
	U = \left(\begin{array}{cccccccccccccc}
	1 \\
	& 1 \\
	&& 1 \\
	&&&&&&&&&&1\\
	&&&&&&&&&-1\\
	&&&&&1\\
	&&-y_2&&&&1 \\
	&&x_1&&&&&1 \\
	y_2 & -y_1 && &&&&&1 \\
	&&&&1 & -y_4\\
	&&&-1&& y_3 \\
	&&&&&&&&&y_3 & y_4 & 1
	\end{array}\right),
	\end{small}
\end{equation}
which makes evident that we are dealing with a $Q$-flux and that the patching conditions are $\beta$ transformations.
We can indeed once more take the coordinates to be compact, provided that whenever we send $y_i \to y_i+1$ we perform a $\beta$ transformation:
\begin{equation}
	\left\{\begin{array}{l}
	y_1 \sim y_1 + 1 \\[2mm]
	\beta^{y_2 \psi_2} = 1
	\end{array}\right., \quad 
	\left\{\begin{array}{l}
	y_2 \sim y_2 + 1 \\[2mm]
	\beta^{y_1 \psi_2} = -1
	\end{array}\right., \quad
	\left\{\begin{array}{l}
	y_3 \sim y_3 + 1 \\[2mm]
	\beta^{y_4 \psi_2} = 1
	\end{array}\right., \quad
	\left\{\begin{array}{l}
	y_4 \sim y_4 + 1 \\[2mm]
	\beta^{y_3 \psi_2} = -1
	\end{array}\right..
\end{equation}

\subsubsection{$x_2 \to 0$ limit}

The limits $x_2 \to 0$ and $x_2 \to  +\infty$ produce mixtures of the previous results.
They produce the same effective theory, with [U(1) $\ltimes T^4]^2$ gauging, but now deriving from either a Q-flux in the $\psi_i$ sector and an asymmetric orbifold in the $\theta_i$ sector, when $x_2 \to 0$, or a Q-flux in the $\theta_i$ sector and an asymmetric orbifold in the $\psi_i$ sector, when $x_2 \to +\infty$.
In detail, the limit $x_2 \to 0$ gives a finite result if we rescale 
\begin{equation}
	y^m \to \left\{x_2 \, \theta_1, \theta_2, \theta_3, x_2\, \psi_1,x_2^2\, \psi_2, \psi_3\right\}
\end{equation}
and 
\begin{equation}
	g \to \frac{g}{\sqrt2\,x_2}.
\end{equation}
The twist matrix has $D$ as above, 
\begin{equation}
	A = \left(\begin{array}{ccc}
	{\mathbb 0}_4  & 0 & 0\\
	0  & 0 & \psi_1\\
	0  & -1 & 0
	\end{array}\right)
\end{equation}
and $R$ as in (\ref{R}), but with $\psi_2 = 0$.

The local geometry is flat in the $\theta_i$ sector and displays a Q-flux in the $\psi_i$ sector. 
These are the corresponding metric
\begin{equation}
	ds^2 =  e^{-\phi/2} \left[d \theta_1^2 + d \theta_2^2  + \theta_1^2 \,d \theta_3^2 +d \psi_1^2 + \frac{1}{1+ \psi_1^2} \left(d \psi_2^2 + \psi_1^2 \,d \psi_3^2\right)\right],
\end{equation}
dilaton
\begin{equation}
	e^\phi = \frac{1}{\sqrt{1+ \psi_1^2}}
\end{equation}
and B-field
\begin{equation}
	B = 2  \frac{\psi_1^2}{1+\psi_1^2}\, d \psi_2 \wedge d \psi_3.
\end{equation}
Note that while the background is not locally flat, it is T-dual to a flat background if we perform a duality along the $\psi_2$ direction.
In flat coordinates, the metric is
\begin{equation}
	\begin{split}
	ds^2 &=e^{-\phi/2}\,\left[ d y_1^2 + dy_2^2 + d \theta_2^2 \right. \\[2mm]
	&\left. +\frac{1}{1+y_3^2+y_4^2}\left( dy_3^4 + dy_4^4 + d \psi_2^2 + (y_3 dy_4 + y_4 dy_3)^2\right)\right],
	\end{split}
\end{equation}
with B-field
\begin{equation}
	B = \frac{2}{1+y_3^2+y_4^2}(y_3 dy_4 - y_4 dy_3)\wedge d \psi_2
\end{equation}
and dilaton
\begin{equation}
	e^\phi = (1+y_3^2+y_4^2)^{-1/2}.
\end{equation}
The non-trivial coordinate identifications are then
\begin{equation}
	 \left\{ \begin{array}{l}
	 \theta_2 \sim \theta_2 + \alpha, \\[2mm]
	 z_L \sim e^{-i \alpha}z_L,\\[2mm]
	 z_R \sim e^{i \alpha}z_R,
	 \end{array}	\right., \qquad
	 \left\{\begin{array}{l}
	y_3 \sim y_3 + 1 \\[2mm]
	\beta^{y_4 \psi_2} = 1
	\end{array}\right., \quad
	\left\{\begin{array}{l}
	y_4 \sim y_4 + 1 \\[2mm]
	\beta^{y_3 \psi_2} = -1
	\end{array}\right..
\end{equation}

\subsubsection{$x_2 \to +\infty$ limit}

The limit $x_2 \to +\infty$ gives 
\begin{equation}
	\begin{split}
	R =& \exp\left[2\left(\psi_2+\frac{\pi}{2}\right)\,C^{46}\right]\,\exp\left[2 \left(\theta_3+\frac{\pi}{2}\right) A^{13}+2\left(\psi_3+\frac{\pi}{2}\right)A^{46}\right] \\[2mm]
	&\, \cdot \exp\left[\pi\left(A^{23}+A^{56}\right)\right]
	\end{split}
\end{equation}
and
\begin{equation}
	A = \left(\begin{array}{cccc}
	0  & 0 & 0 & 0\\
	0  & 0 & \theta_1 & 0\\
	0  & -1 & 0 & 0 \\
	0 & 0 & 0 & {\mathbb 0}_3
	\end{array}\right) 
\end{equation}
and corresponds to the following metric
\begin{equation}
	ds^2 =  e^{-\phi/2} \left[d \theta_1^2 + \frac{1}{1+ \theta_1^2} (d \theta_2^2  + \theta_1^2 \,d \theta_3^2) +d \psi_1^2 + d \psi_2^2 + \psi_1^2 \,d \psi_3^2\right],
\end{equation}
dilaton
\begin{equation}
	e^\phi = \left(1+ \theta_1^2\right)^{-1/2}
\end{equation}
and B-field
\begin{equation}
	B = 2  \frac{\theta_1^2}{1 + \theta_1^2}\, d \theta_2 \wedge d \theta_3.
\end{equation}
The flat space local geometry and its global identifications are analogous to the one presented above, but switching the first 3 and the second 3 coordinates.

\subsubsection{$x_2,x_3 \to 0$ limit}

The combined limit $x_2,x_3 \to 0$ gives again an asymmetric orbifold, with $R$ as in (\ref{R}), but with $\psi_2 = 0$, $A = 0$, and thefore flat metric, vanishing B-field and dilaton.
The corresponding gauge algebra reduces is a single U(1) $\ltimes T^4$ factor.


\subsection{Contractions along $\lambda_i$} 
\label{sub:contractions_along_lambda_i}

The contractions along the $\lambda_i$ generators are trickier.
As expected, sending the corresponding scalar $e \to 0$ or $e \to +\infty$ produces the same gauge group.
For the $e_3$ contractions it is U(1)$^2 \ltimes T^8$, with algebra 
\begin{equation}
	[T_0, T_I] = M_I{}^J T_J, \quad [T_0^\prime, T_I] = N_I{}^J T_J, \quad [T_I, T_J] = 0, \quad [T_0,T_0^\prime] = 0,
\end{equation}
where
\begin{equation}
	M = {\mathbb 1}_4 \otimes i \sigma^2, \quad  	
	N = {\mathbb 1}_2 \otimes \sigma_1 \otimes -i \sigma^2,
\end{equation}
while for the contractions along $e_2$ it is U(1) $\ltimes T^8$, with algebra
\begin{equation}
	[T_0,T_I] = M_I{}^J T_J, \quad [T_I,T_J] = 0,
\end{equation}
where
\begin{equation}
	M = {\mathbb 1}_4 \otimes -i \sigma^2.
\end{equation}
The background geometries are different, though.

\subsubsection{$e_3 \to 0$ limit}

In order to get a finite limit for $e_3 \to 0$ we need to perform the change of coordinates
\begin{equation}
	y^m \to \left\{e_3 \, \theta_1, \frac{1}{\sqrt2}\, (e_3^2\,\psi_2 + \theta_2), \theta_3, e_3\, \psi_1,x_2^2\, \frac{1}{\sqrt2}\, (-e_3^2\,\psi_2 + \theta_2), \psi_3\right\},
\end{equation}
rescaling the coupling constant as
\begin{equation}
	g \to \frac{g}{\sqrt2\, e_3}.
\end{equation}
The resulting twist matrix has
\begin{equation}
	A_2 = \frac{1}{\sqrt2}\, \left(\begin{array}{ccccc}
	{\mathbb 0}_2 & 0 & 0 & 0 & 0  \\
	0 & 0 & 0 & 1 & 0 \\
	0 & 0 & 0 & 0 & 0 \\
	0 & -\theta_1 & 0 & 0 & \psi_1 \\
	0 & 0 & 0 & -1 & 0 
	\end{array}\right),
\end{equation}
and 
\begin{equation}
	\begin{split}
	R =& \exp\left[-\sqrt2 \theta_2 \,C^{13}+(\sqrt2 \psi_2+\pi)\,C^{46}\right]\,\exp\left[2 \left(\theta_3+\frac{\pi}{2}\right) A^{13}+2\left(\psi_3+\frac{\pi}{2}\right)A^{46}\right] \\[2mm]
	&\, \cdot \exp\left[\pi\left(A^{23}+A^{56}\right)\right]\exp\left(\frac{\pi}{2}A^{36}\right).		
	\end{split}
\end{equation}
The corresponding local geometry is described by the metric
\begin{equation}
	\begin{split}
	ds^2 &=  e^{-\phi/2} \left[d \theta_1^2 + d \theta_2^2   +d \psi_1^2 + \frac{2\psi_1^2}{2 + \psi_1^2} d \psi_3^2 \right.\\[2mm]
	&\left. + \frac{1}{2+ \theta_1^2 + \psi_1^2} \left(2 d \psi_2^2 + \theta_1^2 \left(\sqrt{2+ \psi_1^2} \, d \theta_3 + \frac{\psi_1^2}{\sqrt{2+\psi_1^2}}d \psi_3\right)^2 \right) \right],
	\end{split}
\end{equation}
dilaton
\begin{equation}
	e^\phi = \frac{\sqrt2}{\sqrt{2+\theta_1^2 + \psi_1^2}}
\end{equation}
and B-field
\begin{equation}
	B =  e^{2 \phi}\, \sqrt2 \, \left( \theta_1^2 d \theta_3\wedge d \psi_2 + \psi_1^2 \,d \psi_2 \wedge d \psi_3\right).
\end{equation}
This background is not locally flat, though it is T-dual to a flat background if we perform the duality along the $\theta_3$ and $\psi_3$ directions.

Also in this case we have a Q-flux geometry that is evident if we perform the change of coordinates 
\begin{equation}\label{coordchange2}
		y_1 = \theta_1 \sin \theta_3, \quad y_2 = \theta_1 \cos \theta_3, \quad \theta = \frac{\theta_2}{\sqrt2} \quad
		y_3 = \psi_1 \sin \psi_3, \quad y_4 = \psi_1 \cos \psi_3, \quad \psi = \sqrt2\, \psi_2.
\end{equation}
The twist matrix in flat coordinates is
\begin{equation}
	\begin{tiny}
		U =\left(
\begin{array}{cccccccccccc}
 \cos \theta & 0 & 0 & 0 & 0 & 0 & 0 & \sin \theta & 0 & 0 & 0 & 0 \\
 0 & \cos \theta & 0 & 0 & 0 & 0 & -\sin \theta & 0 & 0 & 0 & 0 & 0 \\
 0 & 0 & 1 & 0 & 0 & 1 & 0 & 0 & 0 & 0 & 0 & 0 \\
 0 & 0 & 0 & \sin \theta & 0 & 0 & 0 & 0 & 0 & 0 & \cos \theta & 0 \\
 0 & 0 & 0 & 0 & \sin \theta & 0 & 0 & 0 & 0 & -\cos \theta & 0 & 0 \\
 0 & 0 & -\frac{1}{2} & 0 & 0 & \frac{1}{2} & 0 & 0 & 0 & 0 & 0 & 0 \\
 0 & \sin \theta & -\frac{y_2}{2} & 0 & 0 & \frac{y_2}{2} & \cos \theta & 0 & 0 & 0 & 0 & 0 \\
 -\sin \theta & 0 & \frac{y_1}{2} & 0 & 0 & -\frac{y_1}{2} & 0 & \cos \theta & 0 & 0 & 0 & 0 \\
 0 & 0 & 0 & 0 & 0 & 0 & 0 & 0 & \frac{1}{2} & 0 & 0 & \frac{1}{2} \\
 0 & 0 & \frac{y_4}{2} & 0 & \cos \theta & -\frac{y_4}{2} & 0 & 0 & 0 & \sin \theta & 0 & 0 \\
 0 & 0 & -\frac{y_3}{2} & -\cos \theta & 0 & \frac{y_3}{2} & 0 & 0 & 0 & 0 & \sin \theta & 0 \\
 -y_2 \cos \theta & y_1 \cos \theta & 0 & y_4 \sin \theta & -y_3 \sin \theta & 0 & -y_1 \sin \theta & -y_2 \sin \theta & -1 & y_3 \cos \theta & y_4 \cos \theta & 1 \\
\end{array}
\right),
	\end{tiny}
\end{equation}
which produces the metric 
\begin{equation}
	ds^2 = e^{-\phi/2}\left[2\,d \theta_2^2+ dy_i^2 +\frac{1}{2 + y_i^2} \left(d \psi^2-(y_1 dy_2 - y_2 dy_1 - y_3 dy_4 +y_4 dy_3)^2\right)\right],
\end{equation}
the B-field 
\begin{equation}
	B =  e^{2 \phi}\, \sqrt2 \, \left(-y_1 d y_2 + y_2 d y_1 +y_3 d y_4 - y_4 d y_3\right)\wedge d \psi_2 
\end{equation}
and the dilaton 
\begin{equation}
	e^\phi = \sqrt2 (2 + y_i^2)^{-1/2}.
\end{equation}
We can see that this local geometry corresponds to a Q-flux, because a shift in the $x_i$ coordinates is compensated by a $\beta$ deformation:
\begin{equation}
	\left\{\begin{array}{l}
	y_1 \sim y_1 + 1 \\[2mm]
	\beta^{y_2 \psi_2} = -1
	\end{array}\right., \quad 
	\left\{\begin{array}{l}
	y_2 \sim y_2 + 1 \\[2mm]
	\beta^{y_1 \psi_2} = 1
	\end{array}\right. \quad
	\left\{\begin{array}{l}
	y_3 \sim y_3 + 1 \\[2mm]
	\beta^{y_4 \psi_2} = 1
	\end{array}\right. \quad
	\left\{\begin{array}{l}
	y_4 \sim y_4 + 1 \\[2mm]
	\beta^{y_4 \psi_2} = -1
	\end{array}\right.
\end{equation}
The identification of the remaining coordinate is that of a freely acting orbifold:
\begin{equation}
	\left\{\begin{array}{l}
	\theta \sim \theta + \pi \\[2mm]
	y_i \sim - y_i
	\end{array}\right..
\end{equation}

\subsubsection{$e_3 \to +\infty$ limit}

The limit $e_3 \to +\infty$ produces the same background, only with the exchange of $\theta_2$ and $\psi_2$.

\subsubsection{$e_2 \to 0$ limit}

A different background is obtained if we take the limits of $e_2$ to the boundary.
When $e_2 \to 0$ we get a finite result if we also perform a gauge transformation for the B-field, summarized by the following matrix action on the twist matrix $U$ 
\begin{equation}
	K = \left(\begin{array}{cc}
	 {\mathbb 1}_6 & W \\
	 {\mathbb 0}_6 & {\mathbb 1}_6
 	\end{array}\right), \qquad W_{2}{}^5 = -W_{5}{}^2 = 2,
\end{equation}
together with the coordinate transformation
\begin{equation}
	y^m \to \left\{e_2 \, \theta_1, \theta_2, \theta_3, e_2\, \psi_1,\psi_2, \psi_3\right\}
\end{equation}
and the gauge coupling rescaling
\begin{equation}
	g \to \frac{g}{\sqrt2\,e_2}.
\end{equation}
The outcome is the flat metric 
\begin{equation}
	ds^2 = e^{-\phi/2}\,\left[d \theta_1^2 + 2\,d \theta_2^2   +d \psi_1^2 + 2 \,d \psi_2^2 + \theta_1^2\left(d \theta_3 - d \psi_2\right)^2 + \psi_1^2\left(d \theta_2 + d \psi_3\right)^2\right],
\end{equation}
with constant dilaton
\begin{equation}
	e^\phi = \sqrt2
\end{equation}
and vanishing B-field.
The twist matrix has $A =0$ and interestingly a $D$ that is not diagonal:
\begin{equation}
	D =  \left(\begin{array}{cccccc}
	1 & 0 & 0 & 0 & 0 & 0 \\
	0 & \sqrt2 & 0 & 0 & 0 &  \psi_1 \\
	0 & 0 &  \theta_1 & 0 & 0 & 0 \\
	0 & 0 & 0 & 1 & 0 & 0 \\
	0 & 0 & - \theta_1 & 0 & \sqrt2 & 0 \\
	0 & 0 & 0 & 0 & 0 &  \psi_1 
	\end{array}\right).
\end{equation}
The full matrix is then obtained by acting with the rotation 
\begin{equation}
	\begin{split}
	R =& \exp\left[-2 \theta_2 \,C^{13}+2\left(\psi_2+\frac{\pi}{2}\right)\,C^{46}\right]\,\exp\left[2 \left(\theta_3-\frac{\pi}{2}\right) A^{13}+2\left(\psi_3+\frac{\pi}{2}\right)A^{46}\right] \\[2mm]
	&\, \cdot \exp\left[\pi\left(A^{23}+A^{56}\right)\right]\exp\left(\frac{\pi}{2}\,C^{36}\right).		
	\end{split}
\end{equation}
If we perform the change of coordinates 
\begin{equation}
	\begin{split}
	y_1 &= \theta_1 \, \left(\cos \frac{\psi}{\sqrt2} \sin \theta_3 -\sin \frac{\psi}{\sqrt2} \cos \theta_3 \right), \\[2mm]
	y_2 &= \theta_1 \, \left(\sin \frac{\psi}{\sqrt2} \sin \theta_3 +\cos \frac{\psi}{\sqrt2} \cos \theta_3 \right), \\[2mm]
	y_3 &= \psi_1 \, \left(\cos \frac{\theta}{\sqrt2} \sin \theta_3 +\sin \frac{\theta}{\sqrt2} \cos \theta_3 \right), \\[2mm]
	y_4 &= \psi_1 \, \left(\cos \frac{\theta}{\sqrt2} \cos \theta_3 -\sin \frac{\theta}{\sqrt2} \sin \theta_3 \right), \\[2mm]
	\theta &= \sqrt2 \theta_2, \qquad \psi = \sqrt2 \psi_2,
	\end{split}
\end{equation}
we get a fully flat metric with vanishing dilaton and $B$-field:
\begin{equation}
	ds^2 = d y_1^2 + dy_2^2 + dy_3^4 + dy_4^4+ d \theta_2^2 + d \psi_2^2,
\end{equation}
but with a non-trivial twist matrix 
\begin{equation}
	U = \exp\left[\sqrt2\,\theta \,(C^{12}+A^{45})\right]\,\exp\left[-\sqrt2 \psi ( A^{12}+C^{45})+\pi C^{45}\right]\,\exp\left(\frac{\pi}{2} C^{36}\right) \,.
\end{equation}
The space is therefore a product of an asymmetric and a regular orbifold ($z = y_1 + i\,y_2$, $w = y_3 + i \,y_4$):
\begin{equation}
 \left\{ \begin{array}{l}
	\theta \sim \theta + \alpha, \\[2mm]
 	z_L \sim e^{-i \frac{\alpha}{\sqrt2}}z_L,\\[2mm]
	z_R \sim e^{i \frac{\alpha}{\sqrt2}}z_R, \\[2mm]
	w \sim e^{i \frac{\alpha}{\sqrt2}}w,
 \end{array}\right.
\qquad 
 \left\{ \begin{array}{l}
	\psi \sim \psi + \beta, \\[2mm]
	 w_L \sim i\, e^{-i \frac{\beta}{\sqrt2}}w_L,\\[2mm]
	 w_R \sim -i\,e^{i \frac{\beta}{\sqrt2}}w_R,\\[2mm]
	   z \sim e^{i \frac{\beta}{\sqrt2}}z,
 \end{array}\right.
\end{equation}

\subsubsection{$e_2 \to +\infty$ limit}

Also in this case the limit to the other boundary does not produce anything new, but the same background with the exchange of the $\theta_i$ with $\psi_i$.

\subsubsection{$e_2,e_3 \to 0$ limit}

The double limit $e_2,e_3 \to 0$ produces an effective CSS gauging, related to the flat metric
\begin{equation}
	ds^2 = e^{-\phi/2} \left[d \theta_1^2 + 2 d \theta_2^2  + \theta_1^2 \left(d \theta_3 - \frac{1}{\sqrt2} d \theta_2\right)^2 + d \psi_1^2 + 2 d \psi_2^2 + \psi_1^2 \left(d \psi_3+ \frac{1}{\sqrt2} d \theta_2\right)^2\right],
\end{equation}
with 
\begin{equation}
	e^\phi = \sqrt2
\end{equation}
and vanishing B-field.
The  twist matrix is now determined by 
\begin{equation}
	D =  \left(\begin{array}{cccccc}
	1 & 0 & 0 & 0 & 0 & 0 \\
	0 & \sqrt2 & -\frac{\theta_1}{\sqrt2} & 0 & 0 &  \frac{\psi_1}{\sqrt2} \\
	0 & 0 &  \theta_1 & 0 & 0 & 0 \\
	0 & 0 & 0 & 1 & 0 & 0 \\
	0 & 0 & 0 & 0 & \sqrt2 & 0 \\
	0 & 0 & 0 & 0 & 0 &  \psi_1 
	\end{array}\right),
\end{equation}
$A=0$ and 
\begin{equation}
	\begin{split}
	R =& \exp\left[-\sqrt2 \theta_2 \,C^{13}+2\left(\frac{\psi_2}{\sqrt2}+\frac{\pi}{2}\right)\,C^{46}\right]\,\exp\left[2 \left(\theta_3-\frac{\pi}{2}\right) A^{13}+2\left(\psi_3+\frac{\pi}{2}\right)A^{46}\right]  \\[2mm]
	&\, \cdot \exp\left[\pi\left(A^{23}+A^{56}\right)\right]\exp\left[\frac{\pi}{2}\left(A^{36}+C^{36}\right)\right].		
	\end{split}
\end{equation}
Using again flat coordinates as before
\begin{equation}
	\begin{split}
	y_1 &= \theta_1 \, \left(\cos \frac{\psi}{\sqrt2} \sin \theta_3 -\sin \frac{\psi}{\sqrt2} \cos \theta_3 \right), \\[2mm]
	y_2 &= \theta_1 \, \left(\sin \frac{\psi}{\sqrt2} \sin \theta_3 +\cos \frac{\psi}{\sqrt2} \cos \theta_3 \right), \\[2mm]
	y_3 &= \psi_1 \, \left(\cos \frac{\theta}{\sqrt2} \sin \theta_3 +\sin \frac{\theta}{\sqrt2} \cos \theta_3 \right), \\[2mm]
	y_4 &= \psi_1 \, \left(\cos \frac{\theta}{\sqrt2} \cos \theta_3 -\sin \frac{\theta}{\sqrt2} \sin \theta_3 \right), \\[2mm]
	\theta &= \sqrt2 \theta_2, \qquad \psi = \sqrt2 \psi_2,
	\end{split}
\end{equation}
we get a fully flat metric with vanishing dilaton and $B$-field:
\begin{equation}
	ds^2 = d y_1^2 + dy_2^2 + dy_3^4 + dy_4^4+ d \theta_2^2 + d \psi_2^2,
\end{equation}
The twist matrix is again non-trivial 
\begin{equation}
	U = \exp\left[\theta \,(C^{12}+A^{45})\right]\,\exp\left[-\theta ( A^{12}+C^{45})+\pi C^{45}\right]\,\exp\left(\frac{\pi}{2} (C^{36}+A^{36})\right) \,
\end{equation}
and therefore the space is an asymmetric orbifold ($z = y_1 + i\,y_2$, $w = y_3 + i \,y_4$):
\begin{equation}
 \left\{ \begin{array}{l}
	\theta \sim \theta + \alpha, \\[2mm]
	 z_L \sim z_L,\\[2mm]
	 z_R \sim e^{i \alpha}z_R, \\[2mm]
	 w_L \sim w_L,\\[2mm]
	 w_R \sim e^{-i \alpha}w_R. \\[2mm]
 \end{array}\right.
\end{equation}


\subsection{Mixed contractions} 
\label{sub:mixed_contractions}

To complete the uplift of Table~\ref{tab:contractions}, we need to take mixed contractions between $e_i$ and $x_i$.
The results are always flat metrics with twist matrices corresponding to CSS gaugings.

\subsubsection{$x_3,e_2 \to 0$ limit}

For instance, if we first take the limit $x_3 \to 0$ and then $e_2 \to 0$ we get the local metric
\begin{equation}
	ds^2 = 2\, e^{-\phi/2} \left[d \theta_1^2 + 2 d \theta_2^2  + \theta_1^2 d \theta_3^2  + d \psi_1^2  + 2 d \psi_2^2 + \psi_1^2 \left(d \psi_3+ d \theta_2\right)^2\right],
\end{equation}
with constant dilaton $e^\phi = 4$ and vanishing B-field.
The corresponding twist matrix is determined by 
\begin{equation}
	D =  \left(\begin{array}{cccccc}
	\sqrt2 & 0 & 0 & 0 & 0 & 0 \\
	0 & 2 & 0& 0 & 0 & \sqrt2\, \psi_1 \\
	0 & 0 & \sqrt2 \theta_1 & 0 & 0 & 0 \\
	0 & 0 & 0 & \sqrt2 & 0 & 0 \\
	0 & 0 & 0 & 0 & 2 & 0 \\
	0 & 0 & 0 & 0 & 0 & \sqrt2\, \psi_1 
	\end{array}\right),
\end{equation}
$A=0$ and 
\begin{equation}
	\begin{split}
	R =& \exp\left(-2 \theta_2 \,C^{13}\right)\,\exp\left[2 \left(\theta_3-\frac{\pi}{2}\right) A^{13}+2\left(\psi_3+\frac{\pi}{2}\right)A^{46}\right]  \\[2mm]
	&\, \cdot \exp\left[\pi\left(A^{23}+A^{56}\right)\right]\exp\left(\frac{\pi}{2}\, C^{36}- \pi\,C^{45}\right).		
	\end{split}
\end{equation}
The resulting gauge group is the usual CSS group U(1) $\ltimes T^8$, with algebra
\begin{equation}
	[T_0,T_I] = M_I{}^J T_J, \quad [T_I,T_J] = 0,
\end{equation}
where
\begin{equation}
	M = {\mathbb 1}_4 \otimes -i \sigma^2.
\end{equation}
This flat space is brought to the standard parameterization, with vanishing $B$-field and dilaton, by means of the change of coordinates
\begin{equation}
		\begin{split}
		y_1 &= \sqrt2\,\theta_1 \,\sin \theta_3, \\[2mm]
		y_2 &= \sqrt2\,\theta_1 \,\cos \theta_3, \\[2mm]
		y_3 &= \sqrt2\, \psi_1 \, \left(\cos \theta_2 \sin \psi_3 +\sin \theta_2 \cos \psi_3 \right), \\[2mm]
		y_4 &= \sqrt2\,\psi_1 \, \left(-\sin \theta_2 \sin \psi_3 +\cos \theta_2 \cos \psi_3 \right), \\[2mm]
		\theta &= 2 \theta_2, \qquad \psi = 2 \psi_2.
		\end{split}
\end{equation}
The twist matrix is non-trivial 
\begin{equation}
	U = \exp\left[\theta \,(C^{12}+A^{45})\right]\,\exp\left[ \pi C^{45}\right]\,\exp\left(\frac{\pi}{2} (C^{36})\right) \,
\end{equation}
and leads to the asymmetric orbifold identifications
\begin{equation}
 \left\{ \begin{array}{l}
	\theta \sim \theta + \alpha, \\[2mm]
	z_L \sim e^{-i \alpha/2} z_L,\\[2mm]
	z_R \sim e^{i \alpha/2}z_R, \\[2mm]
	w \sim e^{-i \alpha/2}w
 \end{array}\right.
\end{equation}

\subsubsection{$x_2,x_3,e_3 \to 0$ limit}

The further limit $x_2,x_3 \to 0$ and $e_3 \to 0$ gives again a flat metric in cylindrical coordinates
\begin{equation}\label{metrx5}
	ds^2 = d \theta_1^2 + d \theta_2^2  + \theta_1^2 \,d \theta_3^2 +d \psi_1^2 + d \psi_2^2 + \psi_1^2 \,d \psi_3^2,
\end{equation}
with constant dilaton $e^\phi = 1,$ and vanishing 2-form $B$, for a twist matrix where $A=0$, $D$ as in (\ref{Dorig}) and 
\begin{equation}
	\begin{split}
	R =& \exp\left(-2 \theta_2 \,C^{13}\right)\,\exp\left[2 \left(\theta_3-\frac{\pi}{2}\right) A^{13}+2\left(\psi_3+\frac{\pi}{2}\right)A^{46}\right]  \\[2mm]
	&\, \cdot \exp\left[\pi\left(A^{23}+A^{56}\right)\right]\exp\left(\frac{\pi}{2}\, A^{36}- \pi\,C^{45}\right).	
	\end{split}
\end{equation}
The gauge group is U(1) $\ltimes T^4$, with algebra
\begin{equation}
	[T_0,T_I] = M_I{}^J T_J, \quad [T_I,T_J] = 0,
\end{equation}
where
\begin{equation}
	M = \left(\begin{array}{cc}
	{\mathbb 1}_2 & {\mathbb 0}_2 \\
	{\mathbb 0}_2 &  {\mathbb 0}_2 
	\end{array}\right) \otimes -i \sigma^2.
\end{equation}
This is once more an asymmetric orbifold, once one introduces flat coordinates like in  (\ref{coordchange})
\begin{equation}
	\left\{ \begin{array}{l}
	\theta \sim \theta + \alpha, \\[2mm]
	z_L \sim e^{-i \alpha}z_L,\\[2mm]
	z_R \sim e^{i \alpha}z_R.
	 \end{array}	\right.
\end{equation}

\subsection{Supersymmetry} 
\label{sub:supersymmetry}

Each of the four-dimensional Minkowski vacua uplifted in the previous section preserve a certain amount of supersymmetry, according to Table~\ref{tab:contractions}.
One should be careful, though, of the compatibility of the uplift procedure with the boundary conditions one is imposing to make the background compact.

The 32 supercharges of the ten-dimensional maximal supergravities are encoded in our current DFT setup in terms of two four-dimensional Majorana--Weyl spinors transforming in the $(\mathbf 4,\, \mathbf 1)$ and $(\mathbf 1,\,\mathbf 4)$ of the local symmetry\footnote{For our DFT setup with an external spacetime the easiest way to identify the representations and transformation properties of fermions is to decompose the ones of supersymmetric E$_{7(7)}$ ExFT \cite{Godazgar:2014nqa}.} $\mathrm{SO}(6)_L\times \mathrm{SO(6)}_R \simeq \mathrm{SU}(4)_L\times \mathrm{SU(4)}_R$.

The expression for the Killing spinors for the DFT solutions we have found depends on the choice of $\mathrm{SO}(6)_L\times \mathrm{SO(6)}_R$ gauge.
One possibility is to identify the DFT generalized vielbein with the twist matrix used in the generalized Scherk--Schwarz ansatz.
The twist matrix defines a generalized identity structure on the internal space, such that no local $\mathrm{SO}(6)_L\times \mathrm{SO(6)}_R$ transformations are needed in patching the internal space. 
This holds true also after the freely acting orbifold procedure that yields the $T$-folds of the previous section, as the constant O$(6,6)$ identifications are introduced precisely so that the combination of their action with the coordinate identifications leave the twist matrix invariant.
With this choice, the Killing spinors of the $D=4$ maximal supergravity solutions are lifted as scalar densities as in \cite{Hohm:2014qga}.
Namely, the uplift of the $D=4$ Killing spinors will depend on the internal coordinates only through a power of $\rho(y)$ and will therefore survive the orbifolding if $\rho(y)$ is invariant under the coordinate identifications.
In fact, we have found that this is respected in all our limit geometries, and therefore we can state that the supersymmetries of the $D=4$ solutions summarized in Table 1 uplift to supersymmetries of the asymmetric orbifolds and Q-flux solutions we have found.\footnote{If a non-constant $\rho(y)$ were to jump by constant values under the orbifold action, one could choose to allow for patchings of the fields and Killing spinors involving not just a T-duality but also a trombone rescaling and a shift of the dilaton. We do not encounter this situation here.}

Finally, one may look for other supersymmetries of our DFT solutions that are truncated away in the reduction procedure and therefore did not appear in the four-dimensional models.
One way to approach this is to begin by looking for local solutions of the DFT Killing spinor equations \cite{Coimbra:2011nw,Hohm:2011nu,Jeon:2011sq}, possibly keeping the gauge choice determined by the twist matrix, and then investigating whether any such solutions survive the coordinate identifications.
Clearly, the results might depend on the precise periodicities imposed on these coordinates, and we may then also need to be careful in guaranteeing that the asymmetric orbifold identifications as well as the $\beta$ transformations of our T-fold solutions lie within O$(6,6,\mathbb Z)$.
The amount of residual supersymmetries will depend on the specific conjugacy class of the T-fold monodromies.
A very simple example of this fact is the geometry found for $x_3\to0$, which gives the twist matrix \eqref{x3 to 0 cartesian twist matrix}. 
Clearly, if we impose the periodicity $\theta_2\sim\theta_2+2\pi k$ no asymmetric orbifold identification is needed and the solution is in fact just a torus compactification of ten-dimensional flat space, which is of course fully supersymmetric.



\section{Uplift of general CSS gaugings} 
\label{sec:other_uplifts}

The uplift of the various Minkowski vacua presented in the previous section includes many different CSS gaugings \cite{Cremmer:1979uq} with various supersymmetries.
However, we fail to reproduce the most general class of such gaugings, as one of the mass parameters appearing in these models cannot be tuned when reaching them from limits along the moduli space of the $[\mathrm{SO}(4)\times\mathrm{SO}(2,2)]\ltimes \mathbb{R}^{16}$ model we analyzed \cite{Catino:2013ppa}.\footnote{That mass parameter can be tuned if one starts from the SO$^*(8)$ gauging, or by starting from the other elements of the one-parameter family of $[\mathrm{SO}(4)\times\mathrm{SO}(2,2)]\ltimes \mathbb{R}^{16}$ gaugings found in \cite{Catino:2013ppa} (the parameter being denoted $r_1$ there), whose only known uplift is in terms of section constraint violating twist matrices \cite{Ciceri:2016hup}.}
If we focus on uplifts to eleven dimensions, then a sub-class of the CSS models (depending on three mass parameters) admits standard Scherk--Schwarz uplifts in terms of a twisted $T^7$ \cite{Cremmer:1979uq}.
In this last part of our work we show how to perform the uplift of the general CSS gaugings, depending on all 4 mass parameters introduced in \cite{Cremmer:1979uq}, in terms of a generalized Scherk--Schwarz ansatz.

We start by recalling the interpretation of the 4 mass parameters in terms of fluxes of M-theory.
Following the description in the appendix of \cite{Catino:2013ppa}, 3 of the parameters can be interpreted as torsions on a torus background while the fourth is the flux of the 7-form \emph{and} an additional parameter $\theta_{77}$ that we are going to interpret later:
\begin{equation}\label{CSS fluxes}
	\begin{split}
\omega_{71}{}^2&=-\omega_{72}{}^1=\tilde m_1\,,\\
\omega_{73}{}^4&=-\omega_{74}{}^3=\tilde m_2\,,\\
\omega_{75}{}^6&=-\omega_{75}{}^5=\tilde m_3\,,\\
\theta_{77}&=-g_7 = \tilde m_4\,.
	\end{split}
\end{equation}
The gravitino masses of the gauged supergravity theory, up to an overall factor of some modulus, are given by $m_{1,2,3,4}$ such that 
\begin{equation}
	\begin{split}
\tilde m_1 &= m_1+m_2-m_3-m_4\,,\\
\tilde m_2 &= m_1-m_2+m_3-m_4\,,\\
\tilde m_3 &= m_1-m_2-m_3+m_4\,,\\
\tilde m_4 &= m_1+m_2+m_3+m_4\,.
	\end{split}
\end{equation}
The embedding tensor of $D=4$ maximal supergravity sits in the $\mathbf{912}$ E$_{7(7)}$ representation.
Under the SL(8) subgroup, this splits as $\mathbf{912}\to\mathbf{36}+\mathbf{36}'+\mathbf{420}+\mathbf{420}'$.
The CSS gaugings are parameterized by some components of the $\mathbf{36}'$ and $\mathbf{420}'$ irreps.
Using underlined indices $\underline A,\underline B,...$ for the \textbf{8} of SL(8) we denote such irreps by $\theta_{\underline{AB}}$ and $B^{\underline A}{}_{\underline{BCD}}$ respectively, and the CSS gaugings are defined by \eqref{CSS fluxes} with $\theta_{88}=-g_7$ and $B^{\underline{p} }{}_{\underline{mn}8} = \omega_{\underline{mn} }{}^{\underline{p} }$, $\underline{m} = 1,...,7$.
In particular, the part of the CSS generators that is contained in SL(8) are parameterized by
\begin{equation}
\Theta_{\underline{AB} }{}^{\underline{C} }{}_{\underline{D} } = \delta^{\underline{C} }{}_{[\underline{A} } \theta_{\underline{B} ]\underline{D} } + B^{\underline{C} }{}_{\underline{DAB} }\,.
\end{equation}
When $\tilde m_4=0$ we only have geometric fluxes. 
We can thus uplift to eleven-dimensional supergravity using a standard Scherk--Schwarz ansatz with internal vielbein
\begin{equation}
	\begin{split}
e^1 &= \mathrm{d} y^1 + \tilde m_1\, y^2 \mathrm{d} y^7\,,\quad
e^2 = \mathrm{d} y^2 - \tilde m_1\, y^1 \mathrm{d} y^7\,,\\
e^3 &= \mathrm{d} y^3 +  \tilde m_2\, y^4\mathrm{d}  y^7\,,\quad
e^4 = \mathrm{d} y^4 - \tilde m_2\, y^3 \mathrm{d} y^7\,,\\
e^5 &= \mathrm{d} y^5 + \tilde m_3\, y^6 \mathrm{d} y^7\,,\quad
e^6 = \mathrm{d} y^6 - \tilde m_3\, y^5 \mathrm{d} y^7\,,\\
e^7 &= \mathrm{d} y^7\,. 
	\end{split}
\end{equation}
We then embed the inverse vielbein into SL(8) as (notice that $e=1$)
\begin{equation}
(U)_{\underline{A} }{}^{\underline{B} } = \begin{pmatrix}
e_a{}^m & 0\\
0 & 1
\end{pmatrix}\,,
\end{equation}
and finally obtain the twist matrix
\begin{equation}
(E^{\text{\tiny CSS}}_{\tilde m_1,\, \tilde m_2,\, \tilde m_3})_A{}^M = 
\begin{pmatrix}
(U)_{\underline{AB} }{}^{\underline{CD} } & 0\\
0& (U^{-T})^{\underline{CD}}{}_{\underline{AB} }
\end{pmatrix}\,.
\end{equation}
This twist matrix solves the generalized Scherk--Schwarz condition \eqref{gSS condition} for E$_{7(7)}$ ExFT.\footnote{We follow the conventions of \cite{Hohm:2013uia,Hohm:2014qga}. In particular, $`Y^MN_PQ` = -12 `t_\alpha^MN``t^\alpha_PQ`-\frac12 `\Omega^MN``\Omega_PQ`$.}
The extended internal coordinates are $Y^M$, $M=1,\ldots,56$ which decompose as $(Y^{\underline{AB}},\,Y_{\underline{AB}})$ under SL(8).
The physical internal coordinates are then embedded as
\begin{equation}\label{section choice}
y^{m=1...7} = Y^{m8}\,.
\end{equation}
The interpretation of embedding tensor components in terms of fluxes in equation \eqref{CSS fluxes} assumed this choice of solution of the section constraint.

The parameter $\theta_{77}$ is usually considered to be a locally geometric flux on a torus (see for instance \cite{Aldazabal:2010ef} and more recently \cite{Lust:2017bwq}).
However, notice that the combination of $\theta_{(\underline{mn})}$ flux with seven-form flux can also be interpreted geometrically as curvature of an internal sphere or hyperboloid \cite{Hohm:2014qga}.
In our case, this would be just an $S^1$, so that in fact $\theta_{77}$ together with $g_7$ can have a fully geometric interpretation on a torus.
For instance, CSO(2,0,6) gauged supergravity, corresponding to the CSS model with $\tilde m_{1,2,3}=0,\ \tilde m_4\neq 0$, has been uplifted in \cite{Hohm:2014qga}.
The twist matrix in the fundamental of SL(8) is just a rotation along the 78 plane:
\begin{equation}
E_{\text{\tiny CSO}} = \begin{pmatrix}
1&     &  &                     &  \\
&\ddots&  &                     &  \\
&      &1 &                     &  \\
&      &  &  \cos\tilde m_4 y^7 & \sin\tilde m_4 y^7  \\
&      &  & -\sin\tilde m_4 y^7 & \cos\tilde m_4 y^7  
\end{pmatrix}\,.
\end{equation}
Because the twist matrix is compact, without imposing any periodicity conditions the CSO(2,0,6) supergravity uplifts to eleven-dimensional supergravity with flat internal space.
We can impose arbitrary periodicities for $y^1\ldots y^6$ and the frame is globally well-defined provided $y^7$ has periodicity multiple of $2\pi/\tilde m_4$. 
In this case the $N=0$ Minkowski vacuum of this gauged supergravity lifts to the torus compactification of the fully supersymmetric vacuum of the eleven-dimensional theory. 
If we impose other periodicities to $y^7$, we can regard the solution as a U-fold type geometry analogous to the asymmetric orbifold class of T-folds of the previous section.

Both uplifts discussed above can now be motivated using the general procedure of \cite{Inverso:2017lrz}.
All CSS gauge groups can be written as
\begin{equation}\label{CSS gauge group}
{\rm U}(1)\ltimes {\mathbb R}^{24}\,,
\end{equation}
whose generators are 
\begin{equation}
	\begin{split}
X_{78} &= \tilde m_1 t_{[12]}+\tilde m_2 t_{[34]}+\tilde m_3 t_{[56]}+\tilde m_4 t_{[78]}\,,\\
X_{a7} &= +\tfrac12 \tilde m_{(a)}\epsilon_{ab} t^8{}_b -\tfrac12 \tilde m_4 t^7{}_a\,,
\qquad a=1,...,6\,,  \\
X_{a8} &= -\tfrac12 \tilde m_{(a)}\epsilon_{ab} t^7{}_b -\tfrac12 \tilde m_4 t^8{}_a\,,\\
X^{aa'}&= \tilde m_{(a)} \epsilon_{ab} t^{ba'78}+\tilde m_{(a')}\epsilon_{a'b'} t^{ab'78}\,,
\text{$a$, $a'$ in different couples}\ .
\end{split}
\end{equation}
Here $t^{\underline A}{}_{\underline B}$ are a basis of SL(8) generators, $t_{\underline{AB}}$ generate its SO(8) subgroup and $t_{\underline{ABCD}}$ generate the rest of E$_{7(7)}$, in a basis defined e.g. in \cite{Catino:2013ppa}.
We also write $\tilde m_{(a)} = \tilde m_1,\ \tilde m_2$ or $\tilde m_3$ depending on the couple to which $a$ belongs (12, 34 or 56).
For special values of the masses some of these generators become linearly dependent.
We can still regard the gauge group to be \eqref{CSS gauge group}, simply some of the ${\mathbb R}$ transformations become neutral under U(1) and ungauged, thus becoming a (trivial) central extension of the non-Abelian gauge algebra.\footnote{The fully centrally extended gauge algebra as defined in \eqref{centrally extended gauge algebra} would in fact be even larger, but we do not need it.}

As reviewed earlier, all generalized Scherk--Schwarz uplifts of a gauged supergravity are obtained from coset spaces constructed from the centrally extended version of the gauge group.
The section constraint must then be solved by the projection $\Theta_A{}^{\underline{m}}$ of the embedding tensor on the coset space generators.\footnote{As already stressed in section 4, there is also a second constrain which is redundant in everything we discuss here.}
In our case, we already know that we want our choice of section to be \eqref{section choice}. 
We thus choose
\begin{equation}
{\cal M}_{\text{\tiny internal}} = \frac{ {\rm U}(1)\ltimes{\mathbb R}^{24} }{ {\mathbb R}^{18} }
\end{equation}
where the ${\mathbb R}^{18}$ include all the ones generated by $X^{aa'}$ and an extra ${\mathbb R}^6\subset{\rm SL}(8)$ tailored so that the projection of $\Theta_A{}^\alpha$ onto coset generators is indeed $\Theta_{A}{}^{\underline{m}} = \begin{pmatrix}\delta_{\underline{AB}}^{\underline{m}8}\\0\end{pmatrix}$, as requested.
Such a choice of quotient is always possible, even when some of the ${\mathbb R}$'s become neutral and ungauged, because they are still part of the centrally extended gauge group.
This means that not only we have found an interpretation for the uplifts above, but also we can see that there is no obstruction to uplifting \emph{all} CSS gaugings with arbitrary masses.
The final twist matrix is remarkably simple, being just the product of those described above:
\begin{equation}\label{CSS four param twist matrix}
E^{\text{\tiny CSS}}_{\tilde m_1,\, \tilde m_2,\, \tilde m_3,\, \tilde m_4} =
E_{\rm CSO}{} \cdot E^{\text{\tiny CSS}}_{\tilde m_1,\, \tilde m_2,\, \tilde m_3}  \,.
\end{equation}

We can prove that \eqref{CSS four param twist matrix} is the correct twist by using an observation also exploited in \cite{Inverso:2016eet}.
Let us define the generalized torsion $\mathbb T(\hat E)_{AB}{}^C$ of an arbitrary $y$-dependent matrix $\hat E$ as the coefficients on the right hand side of \eqref{gSS condition}, so that the generalized Scherk--Schwarz condition becomes $\mathbb T(\hat E)_{AB}{}^C = X_{AB}{}^C$.
The generalized torsion associated with \eqref{CSS four param twist matrix} reads
\begin{equation}
	\begin{split}
{\mathbb T}[E^{\text{\tiny CSS}}_{\tilde m_1,\, \tilde m_2,\, \tilde m_3,\, \tilde m_4}] =\ & 
E_{\rm CSO} \cdot {\mathbb T}[E^{\text{\tiny CSS}}_{\tilde m_1,\, \tilde m_2,\, \tilde m_3}]\\ 
&+ {\mathbb T}[\, (E_{\rm CSO})_A{}^M (E_{\rm CSO})_B{}^N  (E^{\text{\tiny CSS}}_{\tilde m_1,\, \tilde m_2,\, \tilde m_3})_N{}^Q \partial_Q   (E_{\rm CSO}^{-1})_M{}^C  \,]\,,
	\end{split}
\end{equation}
but now we notice that ${\mathbb T}[E^{\text{\tiny CSS}}_{\tilde m_1,\, \tilde m_2,\, \tilde m_3}] = X^{\text{\tiny CSS}}_{\tilde m_1,\, \tilde m_2,\, \tilde m_3}$ is invariant under the rotation generator $t_{78}$, that $E_{\rm CSO}$ only depends on $y^7 \equiv Y^{78}$, and finally that $E^{\text{\tiny CSS}}_{\tilde m_1,\, \tilde m_2,\, \tilde m_3}$ leaves $\partial_{78}$ invariant, so that the complete torsion is just the sum of the torsions of $E^{\text{\tiny CSS}}_{\tilde m_1,\, \tilde m_2,\, \tilde m_3}$ and $E_{\rm CSO}$.
The latter two are equal to the embedding tensors of the respective gaugings, so that in total we have (with self-explanatory notation)
\begin{equation}
{\mathbb T}[E^{\text{\tiny CSS}}_{\tilde m_1,\, \tilde m_2,\, \tilde m_3,\, \tilde m_4}] = 
X^{\text{\tiny CSS}}_{\tilde m_1,\, \tilde m_2,\, \tilde m_3} 
+ X^{\rm CSO}_{\tilde m_4} = X^{\text{\tiny CSS}}_{\tilde m_1,\, \tilde m_2,\, \tilde m_3,\, \tilde m_4} \,.
\end{equation}

It is instructive to look at the vector components $K_{\underline{AB}}{}^m \equiv E_{\underline{AB}}{}^{m8}$ of \eqref{CSS four param twist matrix}, and see that they indeed satisfy a U(1) $\ltimes {\mathbb R}^{12}$ algebra relations (the ${\mathbb R}^{12}$ outside of SL(8) are trivially represented), and that these vectors are still non-trivial even when the associated ${\mathbb R}$ generator becomes a central charge.
The non-vanishing vectors are 
\begin{equation}
	\begin{split}
K_{a7} &= \sin(\tilde m_4 y^7) \partial_a\,,\\
K_{a8} &= \cos(\tilde m_4 y^7) \partial_a\,,\\
K_{78} &= \partial_7 \, +\tilde m_1(y^2\partial_1-y^1\partial_2) 
+\tilde m_2(y^4\partial_3-y^3\partial_4)+\tilde m_3(y^6\partial_5-y^5\partial_6)\,.
	\end{split}
\end{equation}
These are the vectors generating the transitive action of the gauge group on the internal space.
It is straightforward to check that, for instance, when $\tilde m_4 =\tilde m_1$ the vector $K_{18}-K_{28}$ becomes central while still being non-vanishing, consistently with the mass dependence of the gauge group structure constants.
Something analogous happens whenever $|\tilde m_4|=|\tilde m_i|$ for at least one $i=1,2,3$, again consistently with the gauge group structure constants.

Finally, we notice that the patching of the $y^{1,...,6}$ coordinates can be taken to be the same as for $\tilde m_4=0$, while the periodicity of $y_7$ determines whether the uplift is globally geometric of of U-fold type.


\section{Outlook} 
\label{sec:outlook}

There are various natural directions of development of this work.
First, while we chose to focus on contractions along the moduli spaces of Minkowski vacua of gauged maximal supergravities, it is worth stressing that the limiting procedure we describe is much more general and can be applied along any direction along the scalar manifold of a gauged supergravity, regardless of whether it corresponds to a modulus of some vacuum solution or not.
This is true for both the contraction procedure in gauged supergravity and for its uplift to a higher dimensional theory.
This means that we could for instance apply our procedure to vacua of maximal supergravity with non-vanishing cosmological constant, also when the original vacuum has no moduli.
Actually, we could even apply our procedure to supergravity theories with no vacua at all, provided we know their uplift.
In these cases one is not guaranteed to obtain a vacuum after the limit, but will generate a new reduction space where one can carry the generalized SS procedure to relate other gauged supergravities to 10 or 11 dimensions.

Another interesting aspect to be explored is the systematic classification of the flat backgrounds like the ones obtained here to answer the question: Which freely acting (asymmetric) orbifolds of superstring theory admit a truncation to gauged supergravity where supersymmetry is spontaneously broken?
One could also fully analyze the string spectrum of the solutions we discussed to understand whether their 4-dimensional supergravity description is a consistent truncation of the full spectrum or also an effective theory in some regime of validity.

Finally, we still lack the higher-dimensional description of the gauged supergravity models of Table~\ref{tab:contractions} with higher rank gauge groups like those of the first column.
It would be interesting to see if a generalization of the procedure described in section \ref{sec:other_uplifts} can be applied to some of these gaugings to obtain a consistent reduction space, with a local geometry that can be described in terms of generalized twist matrices, like in this work.


\bigskip
\section*{Acknowledgments}

\noindent We would like to thank the organizers and participants to the workshop "String Duality and Geometry" in Bariloche, where this work has been first presented.
The work of GD is supported in part by MIUR-PRIN contract 2017CC72MK\_003.
The work of GI was supported by  STFC consolidated grant ST/P000754/1.


\end{document}